\documentclass[5p]{elsarticle}

\usepackage{lineno,hyperref}

\usepackage{graphicx}
\usepackage{xcolor}
\usepackage{amsmath}
\usepackage{tikz}
\usepackage{enumerate}
\usepackage{url}
\usepackage{xcolor}
\usepackage{makecell}
\usepackage{multirow}
\usepackage{bbding}
\usepackage{tikz}
\usepackage{makecell}
\usepackage{eurosym}
\usepackage{amstext} % for \text
\DeclareRobustCommand{\officialeuro}{%
	\ifmmode\expandafter\text\fi
	{\fontencoding{U}\fontfamily{eurosym}\selectfont e}}

\setcellgapes{1pt}

\usepackage{bbding}

\modulolinenumbers[0]

\journal{Elsevier}

\begin{document}

\begin{frontmatter}

\title{A framework to semantify BPMN models using DEMO business transaction pattern}

% Group authors per affiliation:
\author{Sérgio Guerreiro\fnref{myfootnote}}
\address{INESC-ID, Lisbon, Portugal\\
Instituto Superior T\'{e}cnico, University of Lisbon, Lisbon, Portugal}
\ead{sergio.guerreiro@tecnico.ulisboa.pt}

\author{Pedro Sousa}
\address{INESC-ID, Lisbon, Portugal\\
Instituto Superior T\'{e}cnico, University of Lisbon, Lisbon, Portugal}
%\email{pedro.manuel.sousa@tecnico.ulisboa.pt}


\cortext[mycorrespondingauthor]{Corresponding author}


\begin{abstract}

\textbf{Context:}
BPMN is a specification language widely used by industry and researchers for business process modeling and execution. It defines clearly how to articulate its concepts, but do not provide mechanism to represent the semantics of the produced models.

\textbf{Objective:}
This paper addresses the problem of how to improve the expressiveness of BPMN models, proposing a definition for the semantics of a business process within a BPMN model, and improving the completeness of the models in a systematic manner, so that models can describe far more situations with few extra managed complexity.
	
\textbf{Method:}
We conceive a framework based on the business transaction patterns available in the enterprise ontology body of knowledge to prescribe the foundations of semantic BPMN models.
A tool has been developed to automate the framework.
Then, two industrial proof-of-concepts are used to measure its coverage, both positive and negative, and to argue about our proposal's usefulness.
After that, the proposal is compared with others using a systematic literature review.

\textbf{Results:}
A full BPMN pattern is proposed encompassing the happy flow, the declinations, the rejections and the revocations, without adding any new element to the BPMN specification.
A software tool has been developed, and made publicly available, to support the automatic generation of the BPMN models from the proposed patterns.
Our semantified BPMN pattern allowed the identification of a large amount of implicit, and other not implemented, situations in both proof-of-concepts. 

\textbf{Conclusion:}	
It is concluded that the usage of a semantic solution, grounded on a sound pattern, allows the systematic enrichment of the BPMN models with a bounded effort.
Moreover, to simplify the BPMN executable models' implementation, its elements could be classified as implicit, explicit, or not implemented yet.
Finally, related work indicates that this work is demanded, but no full solutions are available.

\end{abstract}


\end{frontmatter}

%\linenumbers

\section{Introduction}
\label{omegaA.introduction}

A business process is a collection of events, activities, and decisions that brings value to the customers of an organization~\citep{dumas2017fundamentals}. 
The use of graphical notations to express business processes is a common approach to achieve a more concise and precise description.
%The modeling of business processes has been gaining more importance given the facilitation offered by a graphical representation when compared with natural languages.
%
The Business Process Model and Notation (BPMN~\cite{bpmn}) is the \emph{de facto} standard used by industry and researchers for business process modeling and execution.
Beyond modeling discussion, the business process models could also be extended to be used as executable ones. In fact, BPMN specification offers the following three sub-classes as an alternative to full modeling conformance: descriptive, analytic and common executable. The common executable is focused on what is needed for executable process models. 

Still, some challenges remain within the scope of BPMN models. 
%The problem addressed in this paper is the existing language gap between the business processes requirements expressed by clients, using their semantics, and the correspondingly models designed using the BPMN language.
This paper addresses the problem of how to improve the expressiveness of BPMN models, and we address two dimensions of this problem
\emph{(i)} defining the semantics of a business process within a BPMN model, 
\emph{(ii)} improving the completeness of the models in a systematic manner, so that models can describe far more situations with few extra managed complexity.

On the one hand, and as corroborated in the literature,
BPMN defines very clearly how to articulate its concepts, but do not provide a semantic for the consequent model. The meaning is usual expressed in the natural language words that are used to name the activities, events, gateways, \emph{etc.}~\cite{ISI:000259894800007,jovst2016empirical}. 
% BPMN do not include any semantic specification to be followed by the modeller~\cite{ISI:000259894800007,jovst2016empirical}. 
Accordingly with~\cite{10.1145.1824795.1824799}, \emph{``a language is considered to be formal if both its syntax and semantics can be precisely defined. When the semantics is formally defined, sentences in the language then have a unique interpretation''}, this property does not hold in BPMN models.
%Moreover, the knowledge related with businesses that is able to act as a communication facilitator, but offer disparate ways to design variability~\cite{10.1145.3041957}.
In fact, it is on the modeler's responsibility side to interpret the meaning of a given BPMN model. 
Conversely, any other modeler could interpret the same model differently, \emph{e.g.}, uBPMN~\cite{yousfi2016ubpmn} extends BPMN with more notations to deal with ubiquitous computing technologies alleging that BPMN do not offer support for this domain.   
The study in~\cite{ISI:000401600000005} corroborates this problem statement and reinforces it when remote environments are imposed, \emph{e.g.}, due to COVID-19 pandemic. 

On the other hand, the business processes expressed by customers usual refers only to the \emph{``happy flow"} and disregards some other forms of exception handling~\cite{ISI:000276032300004}. This option is not acceptable in business process models under scenarios subject to uncertainty~\cite{jimenez2015generating}
Therefore, real business processes implementations usually relies on those incomplete semantic models, and then, this problem is further propagated, and amplified, to the implementation of the correspondingly BPMN executable process models. 

The implications of this problem within organizations are manifold
\emph{(i)} misunderstanding in business process modeling, reflecting in implementation failures, and
\emph{(ii)} inconsistent design in the business process models, reflecting in time consuming tasks for implementation, re-engineering and knowledge dissemination due to an higher learning curve.

To address this problem we propose a framework that enforces the patterns available in Enterprise Ontology~\cite{dietz2020enterprise} onto BPMN models~\footnote{all the models presented in this paper are publicly available at \url{https://github.com/SemantifyingBPMN/DEMO_complete_pattern_in_BPMN}}. The usage of patterns is also corroborated in the literature.~\cite{fellmann2019business} shows that many publications discuss their patterns in isolation, and no embracing pattern exists for the many business process modeling patterns proposals available. 
We stress that our goal is to study solutions to improve the BPMN models expressiveness using knowledge available from other semantic-based approach, and not to change the BPMN language itself. 

The main contributions of this paper are
\emph{(i)} the increase in BPMN models completeness by adding patterns, with a reduced, and thus managed, increase in complexity. This has major impact on executable models maintenance, reuse and testing; 
\emph{(ii)} the standardization of communications between pools according to Enterprise Ontology;
\emph{(iii)} the exemplification of the solution applied to two proof-of-concepts;
\emph{(iv)} an extensive literature review to assess our work with the available knowledge in this domain; and
\emph{(v)} a public tool to facilitate the generation of the semantified BPMN models.

The paper is organized as follows.
Firstly, Section~\ref{omegaA.background} details the background concepts related with BPMN and DEMO.
Afterwards, Section~\ref{omegaA.foundationaltheory} proposes a mapping between the DEMO complete pattern onto BPMN models.
Then, Section~\ref{omegaA.validation} presents two proof-of-concepts of the solution and discusses the obtained results.
After that, in Section~\ref{omegaA.relatedwork} the literature is surveyed to elate our proposal with others.
Finally, Section~\ref{omegaA.conclusions} concludes the paper and identifies future work.

\section{Background}
\label{omegaA.background}

This section presents the background used throughout the paper. Firstly, the Business Process Model and Notation (BPMN) foundations are introduced, then the Design \& Engineering Methodology for Organizations (DEMO) theory and constructs are explained.

\subsection{Business Process Model and Notation}

\begin{figure*}[tbh]
		\centering { \includegraphics[width=0.7\textwidth]{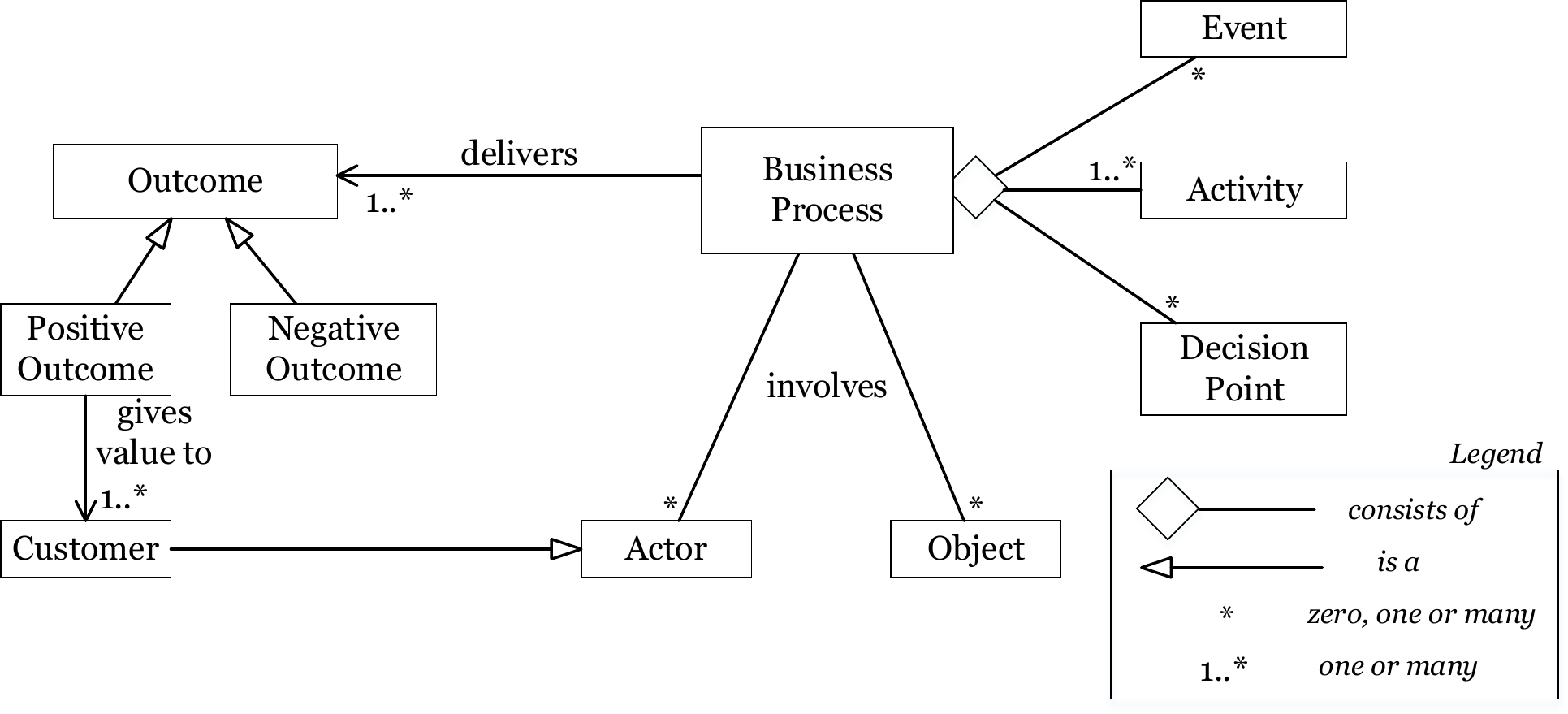} }
       \caption{BPMN metamodel as proposed by~\cite{dumas2017fundamentals}.}
        \label{bpmn.metamodel}
\end{figure*}

A business process is commonly perceived as the sequencing of activities and resources that organizations use to produce value to their clients. The graphical representation of activities, in particular their sequence, is an effective way of modeling processes. The BPMN specification~\cite{dumas2017fundamentals,bpmn} is a graphic process modeling notations used for this purpose. Initially proposed by the Business Process Management Initiative, it is now under the scope of Object Management Group,\footnote{https://www.omg.org/} and the ISO 19510 standard\footnote{https://www.iso.org/standard/62652.html}. BPMN is focused on modeling the articulation activities, resources, flows, gateways, events, messages and data objects that occur in a business process. 

An high-level BPMN meta model proposed by~\cite{dumas2017fundamentals} is depicted in Figure~\ref{bpmn.metamodel}. The most relevant concepts for this paper are the following.
An event which is defined as \emph{``something that ``happens'' during the course of a Process. These Events affect the flow of the Process and usually have a cause or an impact''}~\cite[p.~83]{bpmn}.
An activity which is defined as \emph{``work that is performed within a Business Process'. Activities represent points in a Process flow where work is performed. They are the executable elements of a BPMN Process''}~\cite[p.~151]{bpmn}.
A pool which is defined as \emph{``the graphical representation of a Participant in a Collaboration. A Participant can be a specific PartnerEntity (e.g., a company) or can be a more general PartnerRole (e.g., a buyer, seller, or manufacturer).''}~\cite[p.~112]{bpmn}. This definition is particularly important due to the relation with the forthcoming concept of the DEMO actor role and also due to its separation of a participant by a different pool within the scope of a collaboration.
A lane which is defined as \emph{``a sub-partition within a Process (often within a Pool)''}~\cite[p.~120]{bpmn}.
BPMN supports a detailed specialization of activities, events, and gateways leading to over 100 graphic symbols. Despite such high amount of symbols, BPMN can lead to disparate model interpretations concerning the two core concepts of any BP: resource and activity. In this context, \cite{fickinger2013construct} concluded that BPMN has a level of 51.3\% of overlapping language concepts and lacks state concept to ensure a more sound semantics.

\subsection{Design \& Engineering Methodology for Organizations}

Dietz and Mulder~\cite{dietz2006enterprise} define the notion of Enterprise Ontology (EO) as a set of capabilities to deal with the essential aspects of process-based organizations. 
DEMO aims at understanding the organization using the essential models to represent organizational design and operation. \cite{decosse2014does} show a broad usage of DEMO by the industry, reinforcing the idea that EO is able to capture the essence of the organization while offering abstraction from implementation details. Among other aspects, DEMO prescribes a pattern for the communication and production of acts and facts that occur between actors in the scope of a business transaction. 
A business transaction is defined in~\cite{dietz2006enterprise} as: \emph{``a sequence of coordination-acts\footnote{Accordingly with~\cite{dietz2006enterprise}, the coordination acts (or C-acts for short) are communicative acts in Habermas' category of regulativa~\cite{habermas1984theory}.}/facts\footnote{A facts is considered to exist in the objective world.}, within the transaction pattern, concerning some product. It involves two actors, one in the role of initiator and one in the role of executor. An actor is a subject in filling an actor role''}.
On industrial implementations this pattern is used as an element within a network of transactions. Therefore, to enforce these principles it is of core importance to understand its structure and dynamics.

%\begin{figure}[htbp]
%		\centering { \includegraphics[width=0.3\textwidth]{./images/DEMOconstructs.pdf}	}
%       \caption{Constructs for representing DEMO business transactions (from source~\cite{dietz2006enterprise}). }
%        \label{demo.constructs}
%\end{figure}

%Figure~\ref{demo.constructs} depicts the DEMO constructs for representing business transactions.
%Starting in the top left part of the figure, an elementary actor is represented by a white box, while a composite actor (a network of transactions and other actors inside) is represented by a grey box. 
%A business transaction type is represented by a circle with a diamond inside, and inherits the previous introduced DEMO standard pattern of a transaction definition.
%Next, in the second row of the figure, the boundary of an organization is represented by a grey line where all the business transaction's types and actor roles are designed inside it.
%The actor role executor is represented with an executor link connected to the transaction type. Similarly, the actor role initiator is represented with the initiator link.

\begin{figure}[tbhp]
		\centering
		\scalebox{0.45}{\includegraphics{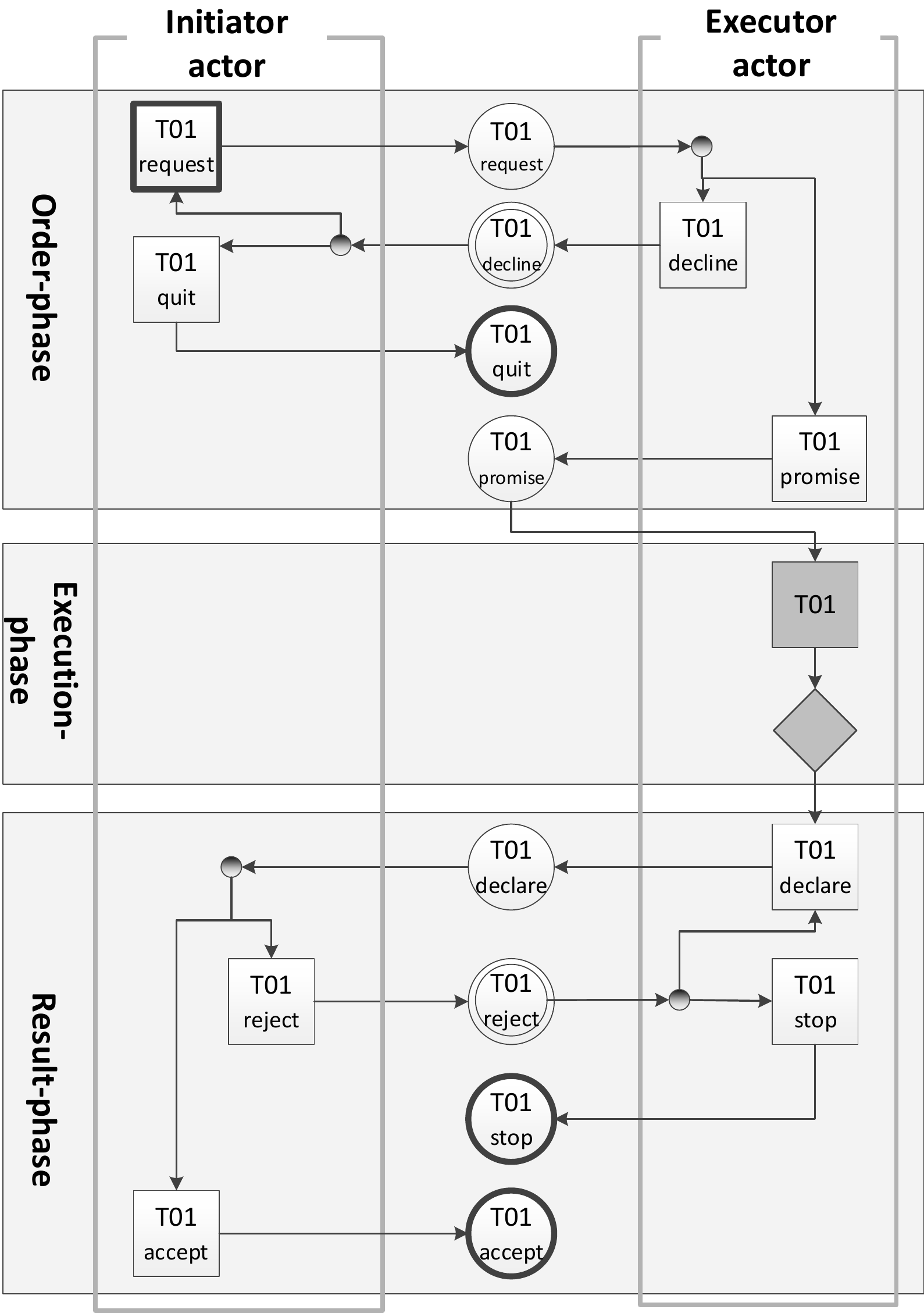}}
        \caption{DEMO standard pattern of a transaction between two actors with separation between communication and production acts (from source~\cite{dietz2006enterprise}).}
        \label{demo.pattern}
\end{figure}

Similarly with other approaches~\cite{sinz2018short,ferstl1998modeling,bork2013bridging}, a DEMO business transaction model has two distinct worlds: \emph{(i)} transition space and \emph{(ii)} state space. DEMO's transition space is grounded in a theory named $\Psi$-theory (PSI), where the standard pattern of a transaction includes two distinct actor roles: the \emph{Initiator} and the \emph{Executor}. These two actor roles are considered the minimal conditions to perform communication within a business transaction: the actor role that requires a new product and the actor role that is able to offer the new product. The transaction pattern is performed by a sequence of coordination and production acts that leads to the production of the new production fact. In detail, it encompasses: \emph{(i)} order phase that involves the acts of \emph{request}, \emph{promise}, \emph{decline} and \emph{quit}; \emph{(ii)} execution phase that includes the production act of the new fact itself; and \emph{(iii)} result phase that includes the acts of \emph{declare}, \emph{reject}, \emph{stop} and \emph{accept}. Firstly, when a Customer desires a new product, he requests it. After the request for the production, a promise to produce is delivered by the Producer. Then, after the production, the Producer declares that the production is available. Finally, the Customer accepts the new fact produced. DEMO's basic transaction pattern aims at specifying the transition space of a system that is given by a set of allowable sequences of transitions (see Figure~\ref{demo.pattern}). 

\subsubsection{Declination and rejection}

Besides the sequence of request, promise, declare and accept (also named as the happy path), \emph{cf.} Figure~\ref{demo.pattern}, the \emph{Coordination-acts} of decline and reject are negative options that actor roles can take during the course of a business transaction execution. A \emph{decline} corresponds to an impossibility of promising the production of a new product, by the executor role, \emph{e.g.}, due to lack of raw-material stock. In that situation it is a decision of the initiator actor to finish the business transaction communicating with a \emph{quit}, or insisting with the \emph{request}. This conversation could drive to a deadlock situation. The deadlock is avoided with common sense between the actors.

Symmetrically, a \emph{reject} could be issued by an initiator role after the new product had been \emph{declared}, \emph{e.g.}, if the declared product is defective. In that situation, the executor role decides if the rejection is acceptable, and if so, a \emph{stop} is emitted. Otherwise, a \emph{declare} is re-emitted to the initiator. The difference to the declination situation is that, now, the new product already exists in the world, and its production involved costs. Therefore, the cost has to be assigned to one of the two actors involved. Usually, if the cost is incurred by the executor actor then a \emph{stop} is issued and the business transaction execution ends.

\subsubsection{Revocation}

DEMO complete transaction pattern also includes the possibility to cancel any of the previous executed act. A revoking act is the consequence of \emph{a posteriori} dissatisfaction, regret, \emph{etc.}, about a fact or act occurred in the past. Moreover, any revocation event can occur at any time during the execution of a business transaction. 
Four different revocation events could occur: 
\emph{revoke a request} - when the initiator actor abort his/her initial request of a product, \emph{e.g.}, I do not want a pizza anymore;
\emph{revoke a promise} - when the executor actor realizes that has no conditions to proceed with a previous promise, \emph{e.g.}, the no longer have ingredients to produce the promised pizza;
\emph{revoke a declare} - when the executor role calls off the declared product, \emph{e.g.}, return me the delivered pizza because that is not yours; and 
\emph{revoke a accept} - when the initiator actor realizes that the previously accepted product is not in good conditions, \emph{e.g.}, reject the delivered pizza because there is a fly on it.
The consequence of a revocation event is one of the following:
\emph{(i) allow}, when the initiator allows a \emph{revoke promise} or a \emph{revoke declare}, or when the executor allows a \emph{revoke request} or a \emph{revoke accept};
\emph{(ii) refuse}, when the initiator refuses a \emph{revoke promise} or a \emph{revoke declare}, or when the executor refuses a \emph{revoke request} or a \emph{revoke accept}. In the case of a \emph{refuse} the revocation can always be re-triggered.

\begin{figure*}[tbhp]
		\centering
		\includegraphics[width=\textwidth]{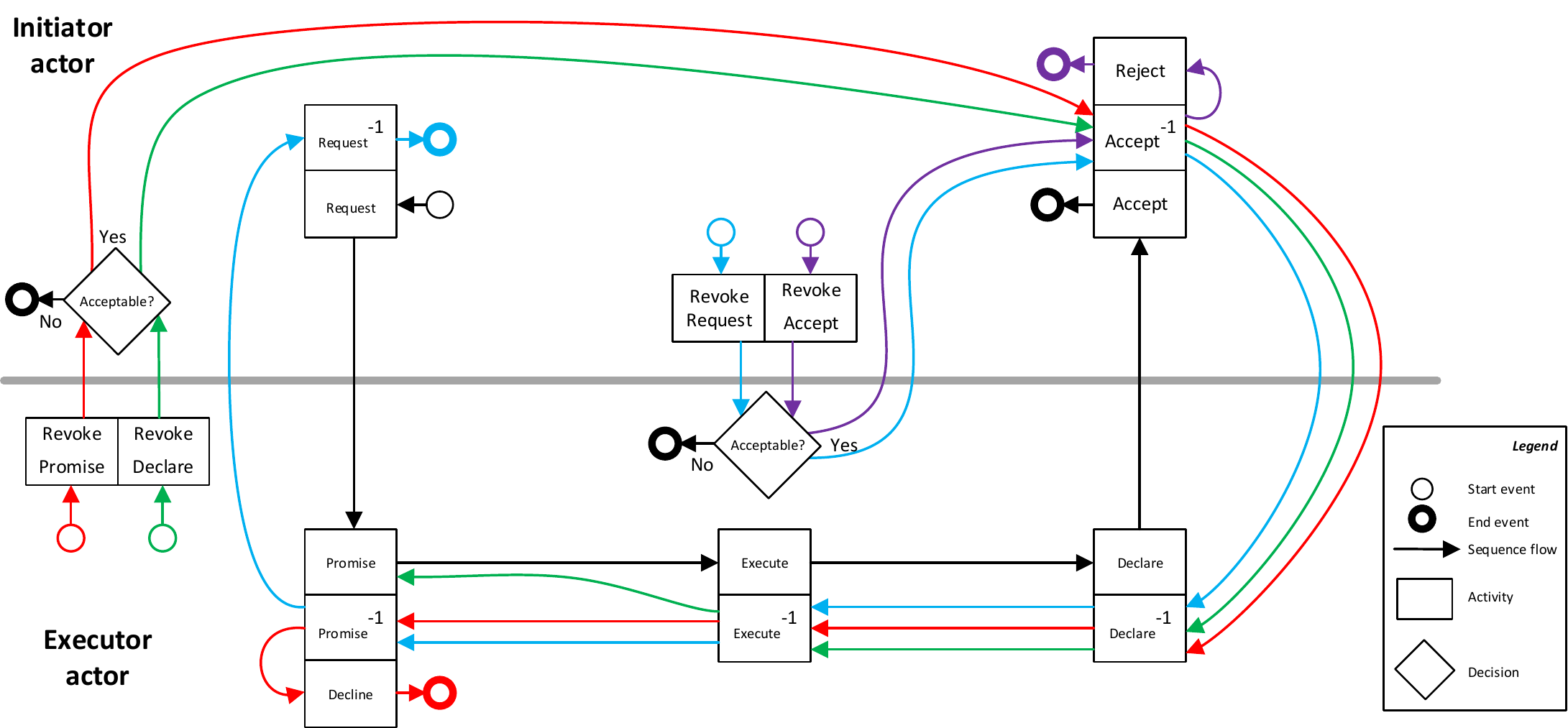}
        \caption{Pseudo-BPMN diagram for the revocation pattern of DEMO standard pattern of a transaction between two actors. Revoke request represented in \textcolor{blue}{blue}, revoke promise represented in \textcolor{red}{red}, revoke declare represented in \textcolor{green}{green} and revoke accept represented in \textcolor{violet}{violet}. Where $^{-1}$ represents the rollback of a transaction step.}
        \label{generic.revocation.pattern}
\end{figure*}

To better visualize this explanation, Figure~\ref{generic.revocation.pattern} depicts the flow that is executed for each one of these four revocation events distinguished by different colors. In general, each revocation encompasses the following sequential phases:
\emph{(i)} first actor role triggers revocation and sent it to the second actor role,
\emph{(ii)} second actor role decides to allow or refuse the revocation,
\emph{(iii)} if allowed then all the previous executed transaction steps in both actor role are rollbacked\footnote{corresponding to the inverse execution of each transaction step} and the transaction is re-positioned in the correspondingly target transaction step, and 
\emph{(iv)} otherwise, refuse is emitted, revocation ends and business transaction proceeds as initially.
Whenever a revocation event is allowed, the recovery flow to be executed depends on the specific event that has been issued. 
In detail, for each revocation, and with increasingly complexity. 
Regarding the \emph{revoke accept}, the rollback of \emph{accept} is performed and then \emph{reject} is emitted. In this situation, the business transaction is relocated in a state similar to a rejection of a \emph{declare}.
Regarding the \emph{revoke declare}, the rollback of \emph{accept}, \emph{declare} and \emph{execute} is performed and then a \emph{promise} re-emitted. In this situation, the initiator knows that his/her request will be re-executed and re-notified by a \emph{declare} with production finish.
Regarding the \emph{revoke promise}, the rollback of \emph{accept}, \emph{declare}, \emph{execute} and \emph{promise} is performed and then a \emph{decline} is emitted. In this situation, the business transaction is relocated in a state similar to a decline of a \emph{request}.
Finally, regarding the \emph{revoke request}, the rollback of all acts need to be performed: \emph{accept}, \emph{declare}, \emph{execute}, \emph{promise} and \emph{request}. In this situation, the business transactions is fully rollbacked. 
As could be verified, revocation complexity increases when we cancel the steps closer to the beginning of the transaction.
	
As a summary, the full list of \emph{coordination-acts} are the following: \emph{request}, \emph{promise}, \emph{declare}, \emph{accept}, \emph{decline}, \emph{reject}, \emph{revoke request}, \emph{revoke promise}, \emph{revoke declare}, \emph{revoke accept}, \emph{allow}, \emph{refuse} and \emph{stop}.
The only \emph{production-acts} is the \emph{execute}.

\section{Mapping DEMO complete pattern onto BPMN models}
\label{omegaA.foundationaltheory}

This section is grounded in the previous concepts introduced on section~\ref{omegaA.background}, and to better explain the conceptualization beneath DEMO $\Psi$-theory an example is used. Therefore the DEMO principles are used but expressed in BPMN notation. The objective is to combine a well-known notation (BPMN) with a comprehensive business process semantics definition (DEMO).

\begin{figure}[htbp]
		\centering { \includegraphics[width=0.3\textwidth]{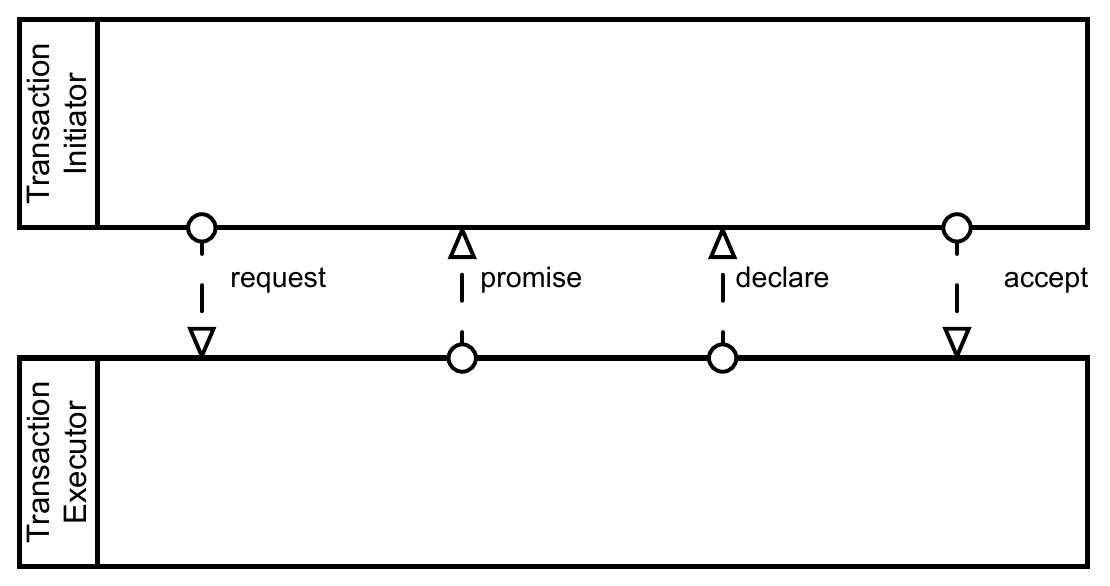}	}
       \caption{The DEMO standard pattern of a transaction between two actors represented as two BPMN public pools.}
        \label{demo.public}
\end{figure}

\begin{figure*}[htbp]
		\centering { \includegraphics[width=1.0\textwidth]{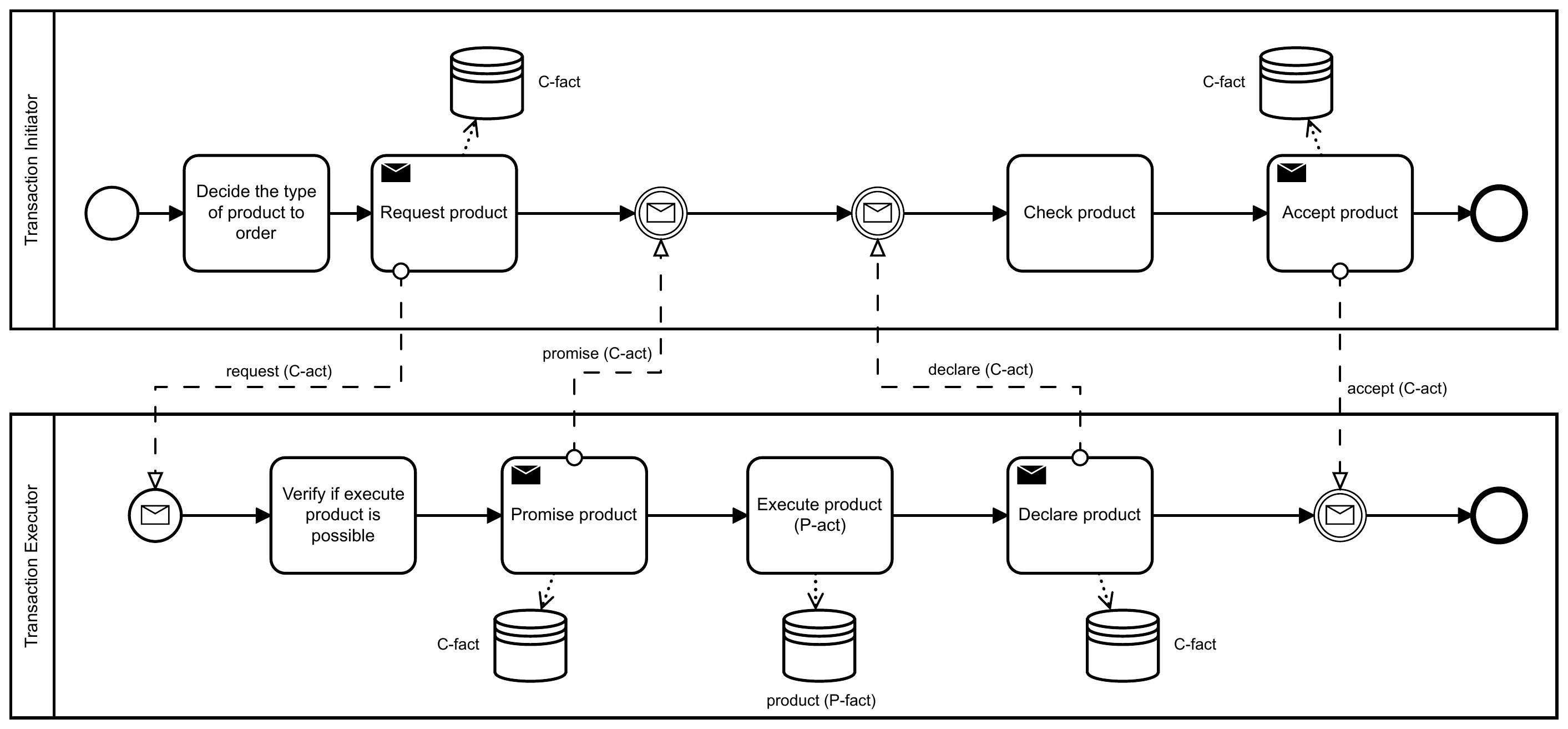}	}
       \caption{The ``happy flow'' of the DEMO standard pattern of a transaction between two actors with separation between communication and production acts represented in BPMN.}
        \label{demo.pattern2}
\end{figure*}

Considering a single business transaction of \emph{producing a product}; at least two business actors need to be considered: the transaction initiator (TI) and the transaction executor (TE). Considering that both actors are located on two different organizations, then two BPMN pools are considered (\emph{cf.} Figure~\ref{demo.public}). Withal, this option is also aligned with the BPMN pool definition. The first initiates the transaction and the second executes it. Moreover, the second only starts production in reaction to an explicit request from the first one.
When TI decides to request a product, the task of \emph{Request product} is executed resulting in \emph{(i)} the emission of a communication act (C-act) with the purpose of \emph{Requesting a product} to the TE, and \emph{(ii)} the creation of a new communication fact (C-fact) in the world. This double outcome is repeated in all the business transaction steps, except on the \emph{Execute product} step where no C-act is expected but a P-act is. Figure~\ref{demo.pattern2} expands all the business transaction steps included in a DEMO standard pattern, clarifying how the sequence of states originate a new P-fact in the world. 
Whenever an actor role is expecting a C-act from the other, then a BPMN interrupting message intermediate event is used.
All the business transactions respect this pattern, even when some acts or facts are not observable. In that situation, the acts or facts are considered implicit in the execution of the business transaction.

\subsection{Declination and rejection}
\begin{figure*}[htbp]
		\centering { \includegraphics[width=1.0\textwidth]{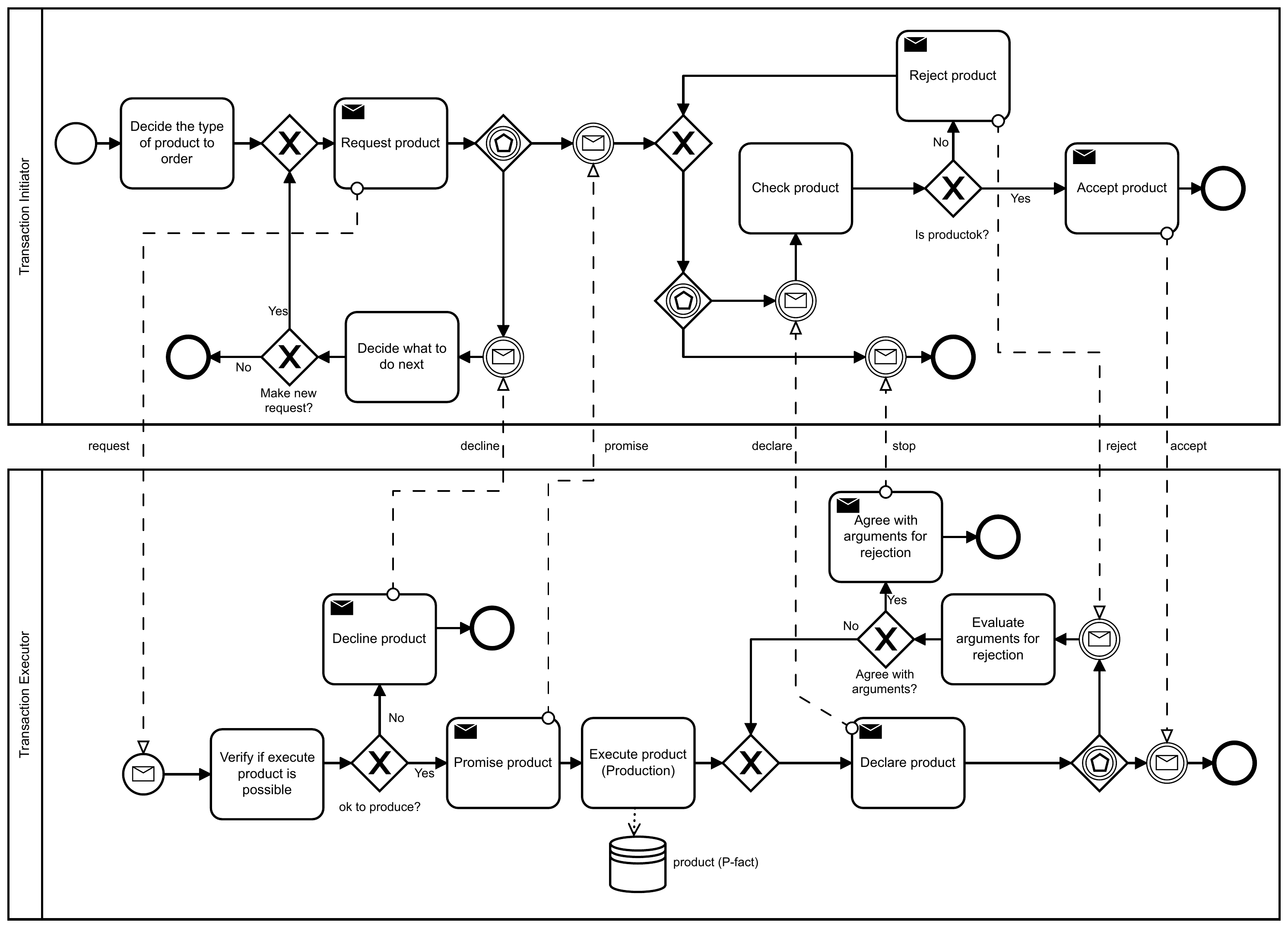}	}
       \caption{The DEMO standard pattern of a transaction between two actors with separation between communication and production acts, and presenting the disagreements, represented in BPMN.}
        \label{demo.pattern3}
\end{figure*}

A more detailed pattern is presented in Figure~\ref{demo.pattern3}. This time, the disagreements between actors are considered, namely declining to produce a product and the rejection of a produced product. If a business actor disagrees with another, then a decision point is achieved. In case of declination, it is up to the TI to issue a new request; on the contrary, in case of rejection, the TE needs to evaluate the rejection arguments before deciding between stopping or re-declaring the product. Both situations can lead to a deadlock situation; it is the context of social norms that prevents a situation of an eternal deadlock. 
On the one hand, the C-act decline corresponds to an impossibility of producing a product (\emph{e.g.}, due to stock shortage) and triggers the TE end of process flow. On the other hand, TI upon the reception of a declination could decide to emit a new request or also to stop his/her process flow.
The rejection C-act is symmetric with decline. Whenever a TI receives a C-act declare then a product check is performed. If TI do not agree with the delivered product then a C-act of rejection is emitted. Understanding that P-fact already exists in the world, is on the TE responsibility to decide if the rejection is valid or not. If so, a stop C-act is emitted and the TI and TE process flow ends. Otherwise, the declare C-act is re-emitted and the loop is restarted. Again, a possible deadlock could occur, and in the limit can only be solved using litigation.   
The synchronization between both BPMN pools is assured using the event-based gateway. This gateway permits the process sequence to evolve accordingly with the act communicated by the other pool (\emph{cf.} its definition on~\cite[p.~297]{bpmn} \emph{``Basically, the decision is made by another Participant, based on data that is not visible to Process, thus, requiring the use of the Event-Based Gateway''}.
Whenever any event defined in an event-based gateway is triggered, then the outgoing path will be used and all the remaining paths will no longer be valid~\cite[p.~298]{bpmn}. 
\subsection{Revocations}
\begin{figure*}[tbhp]
		\centering
		\includegraphics[width=\textwidth]{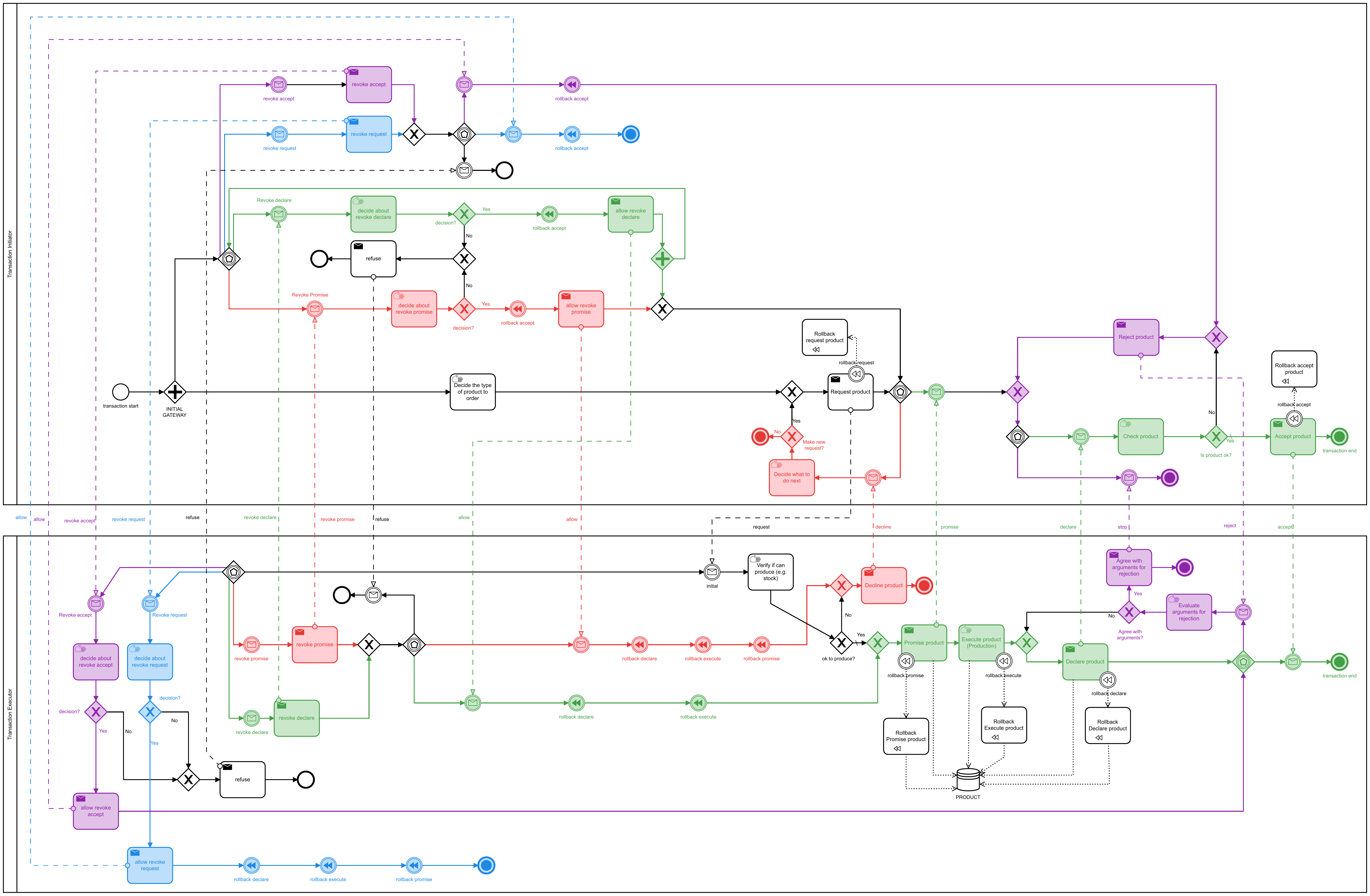}		
        \caption{DEMO complete pattern of a transaction between two actors with separation between communication and production acts and supporting all the: request, accept, promise and declare revocations. Represented in BPMN.}
        \label{demo.pattern.request.accept.promise.decline.revocation}
\end{figure*}

Whenever an actor role, within the scope of any business transaction, need to revoke a previous executed transaction step, the revocation pattern is triggered.
As defined in Figure~\ref{generic.revocation.pattern}, DEMO complete transaction pattern offers the four following revocations differentiated by color: \emph{revoke a request}, \emph{revoke a promise}, \emph{revoke a declare} and \emph{revoke a accept}. 
Figure~\ref{demo.pattern.request.accept.promise.decline.revocation} depicts all the possible revocations using the same colors: \textcolor{blue}{revoke request}, \textcolor{red}{revoke promise}, \textcolor{green}{revoke declare} and \textcolor{violet}{revoke accept}.
The mapping of DEMO revocations onto BPMN requires the addition of two BPMN elements to the model presented in Figure~\ref{demo.pattern3}: the multiple start event and the compensation event.

Multiple start events are necessary because while the normal flow of a process executes a revocation act could happen. 
In other words, the triggering of the transaction start or one of the four revocations are independent events.
For that reason, in TI, a start event is placed before a parallel gateway (named in the model as \emph{INITIAL GATEWAY}) that triggers two distinct branches: the normal flow corresponding to the DEMO transaction pattern and an event-based gateway catching all the message events related with the revocations. The same applies to TE, except that the start event is replaced by a message start event corresponding to a communication received from TI. This solution allows the triggering of any revocation anytime. Then, a communication with the other actor role is established to assert if the revocation could be acceptable, \emph{e.g.}, a \emph{request revoke} is triggered in TI, sent to TE and decided by TE. If TE agrees, then communicates positively with TI (which by its turn will rollback the transaction steps of its competence, namely, rolling back the accept P-act) and then rollbacks the P-acts of declare, execute and promise. In this \emph{request revoke} example, both pools also trigger the terminate event to end all running process instances. In the opposite, if TE do not agree with the revocation, then a refuse is communicated to TI and the normal flow continues. All the remaining revocations are modeled in Figure~\ref{demo.pattern3} in alignment with Figure~\ref{generic.revocation.pattern}.

Moreover, when a revocation of a non-started transaction is triggered it is on the other actor's responsibility to reject it. For instance, when a TI revokes an accept of a transaction that has not been even started. This situation could be considered as a deception attempt and rejected.

Compensation events are the native mechanism in BPMN to trigger a new compensation flow, which is defined as: \emph{``the flow that defines the set of activities that are performed while the transaction is being rolled back to compensate for activities that were performed during the normal flow of the process''}~\cite[p.~500]{bpmn}. Our solution relies on compensation events classified with the name of the revocation act to be performed. 

Finally, the possibility of revocation recurrence is studied. The recurrence of revocations only make sense in a \emph{revoke declare} situation where the transaction is reverted to the promise step and resumed from there.
For the remaining situations the following behavior is expected.
In the case of a \emph{revoke request} the transaction is ended.
In the case of a \emph{revoke promise} the transaction is declined.
In the case of a \emph{revoke accept} the transaction is rejected.

%\subsection{Implicit and explicit communication acts}

\subsection{Patterns for transaction composition}
In real environments, the business transactions are defined within a network of actors and transactions, \emph{e.g.}, a \emph{Payment} transaction succeeds a \emph{Production} transaction. Applied scenarios of these situations are detailed in Section~\ref{omegaA.validation}.

\begin{figure}[htbp]
		\centering { \includegraphics[width=0.45\textwidth]{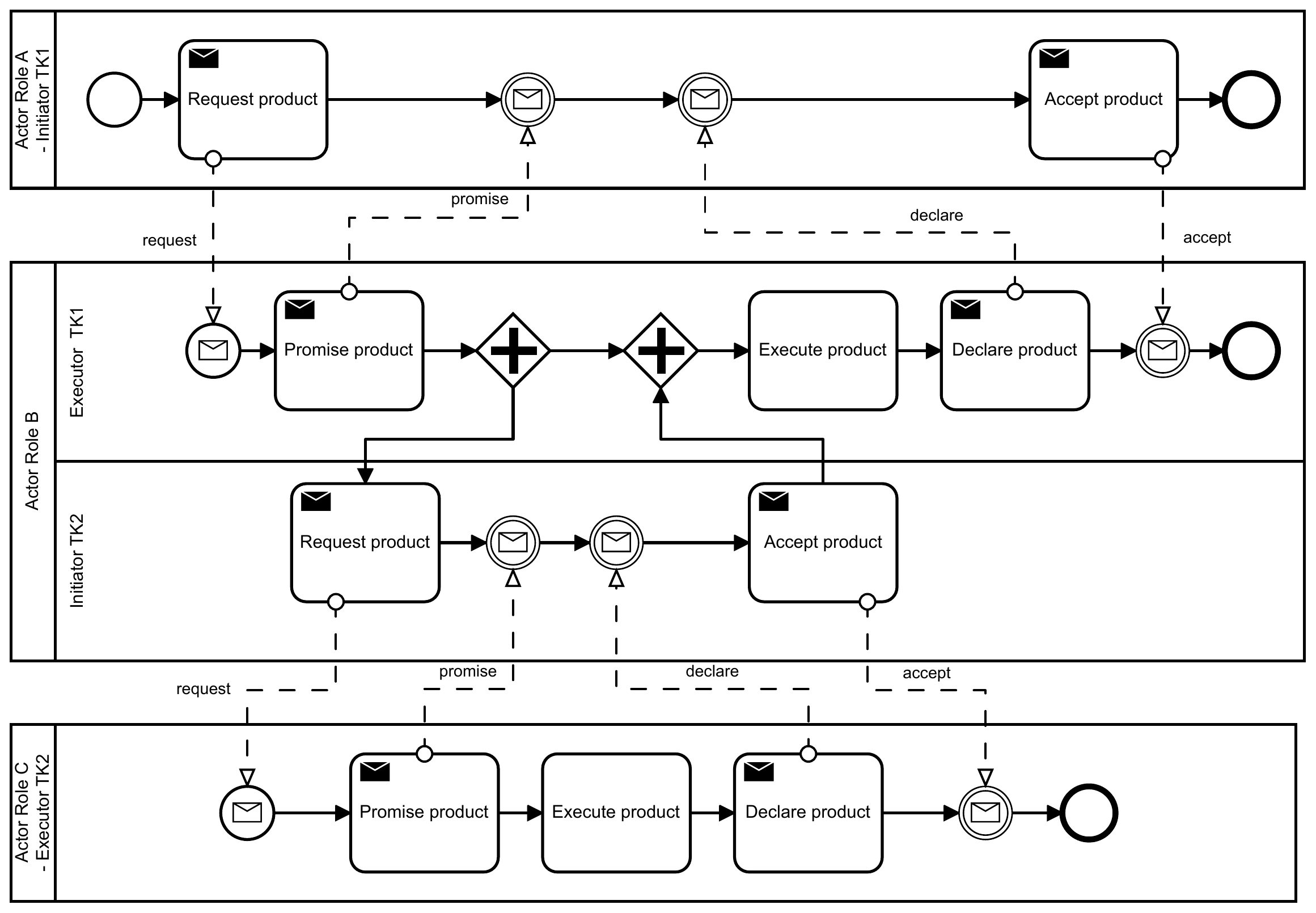}	}
       \caption{Dependency between transactions: request after promise (RaP) composition pattern concerning 2 business transactions.}
        \label{demo.patternRaP2tx}
\end{figure}

\begin{figure}[htbp]
		\centering { \includegraphics[width=0.45\textwidth]{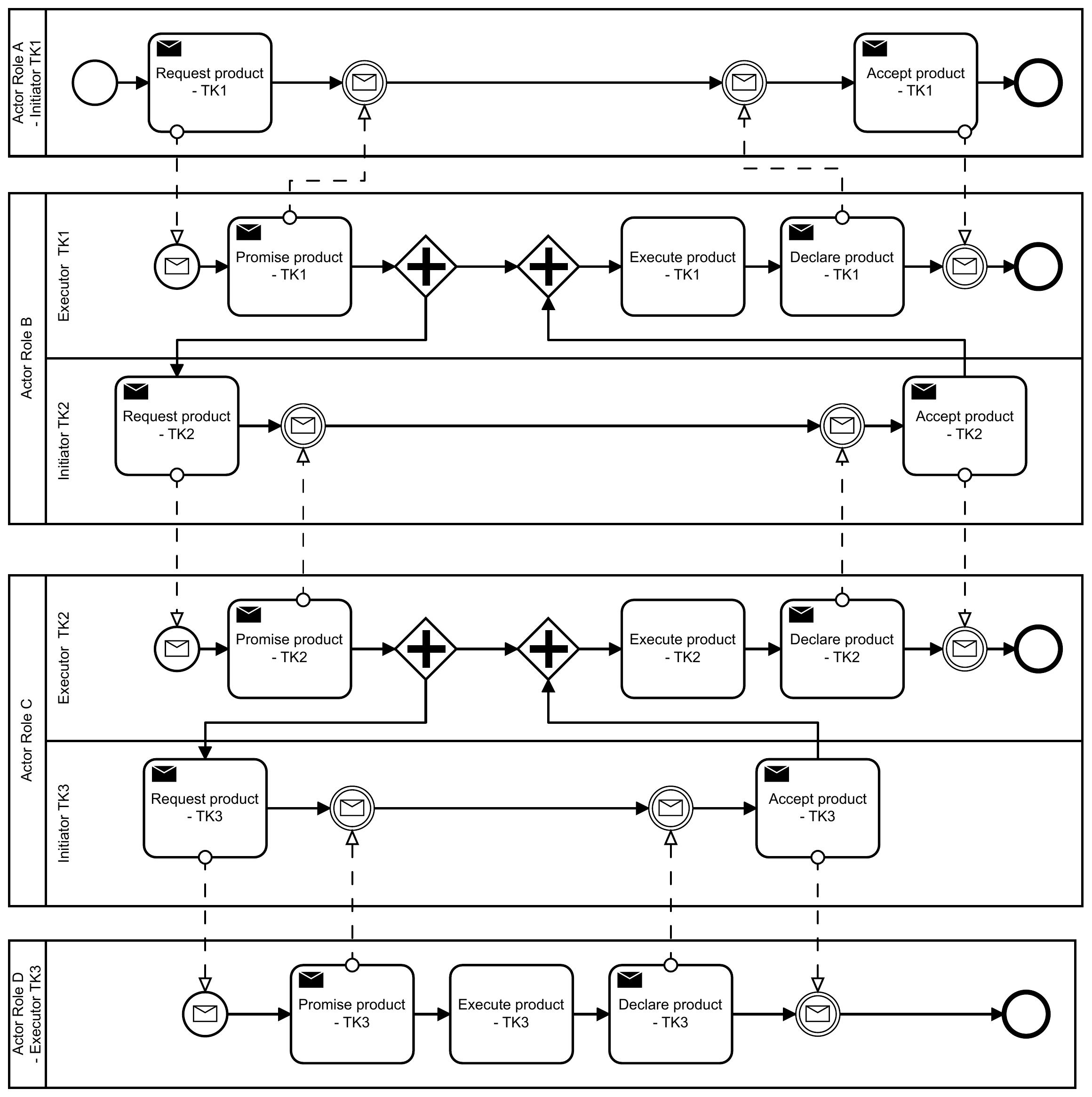}	}
      \caption{Dependency between transactions: request after promise (RaP) composition pattern concerning 3 business transactions.}
        \label{demo.patternRaP3tx}
\end{figure}

\begin{figure}[htbp]
		\centering { \includegraphics[width=0.45\textwidth]{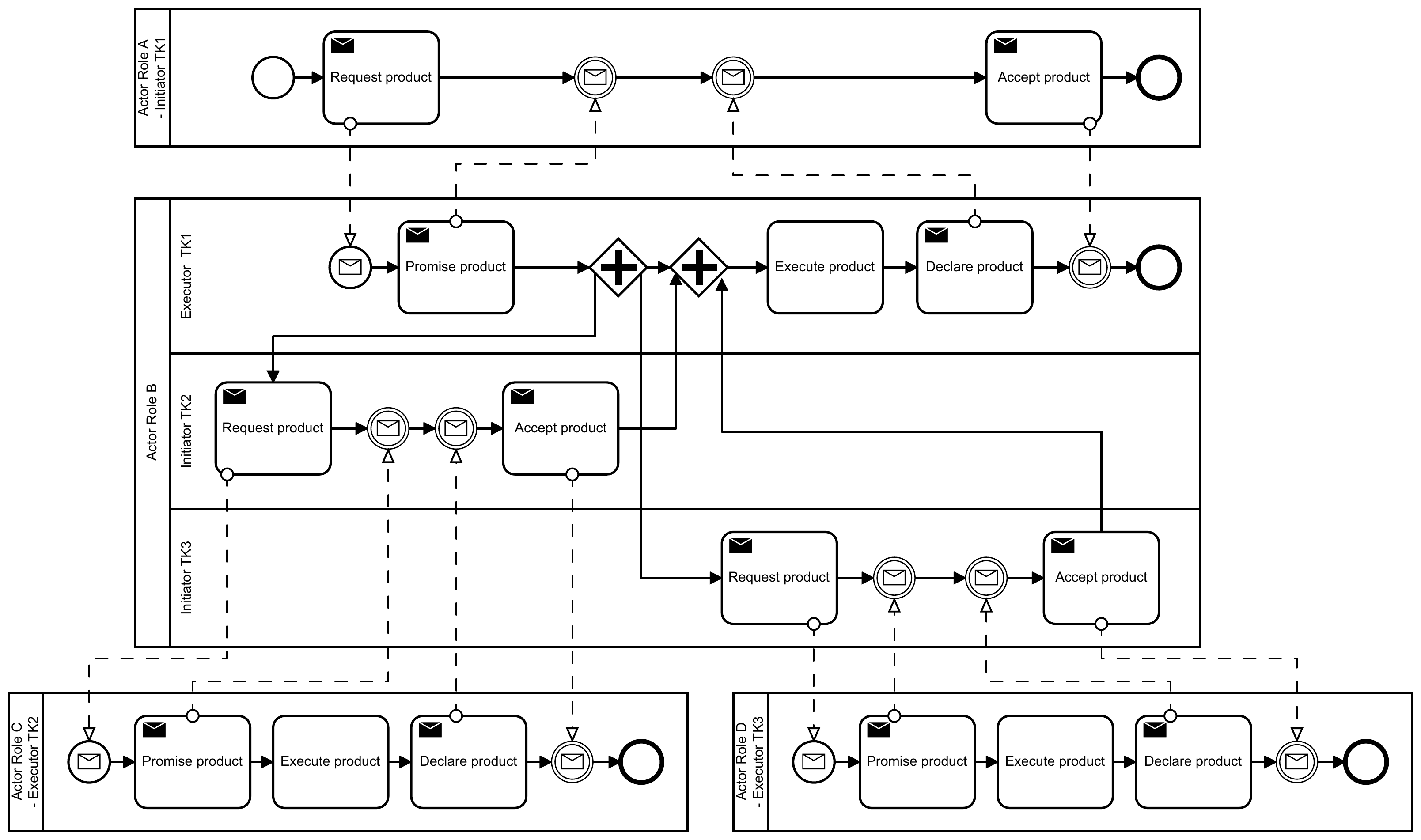}	}
      \caption{Dependency between transactions: request after promise (RaP) composition pattern concerning 3 business transactions where TK2 and TK3 is performed simultaneously.}
        \label{demo.patternRaP3txsimul}
\end{figure}

\begin{figure}[htbp]
		\centering { \includegraphics[width=0.45\textwidth]{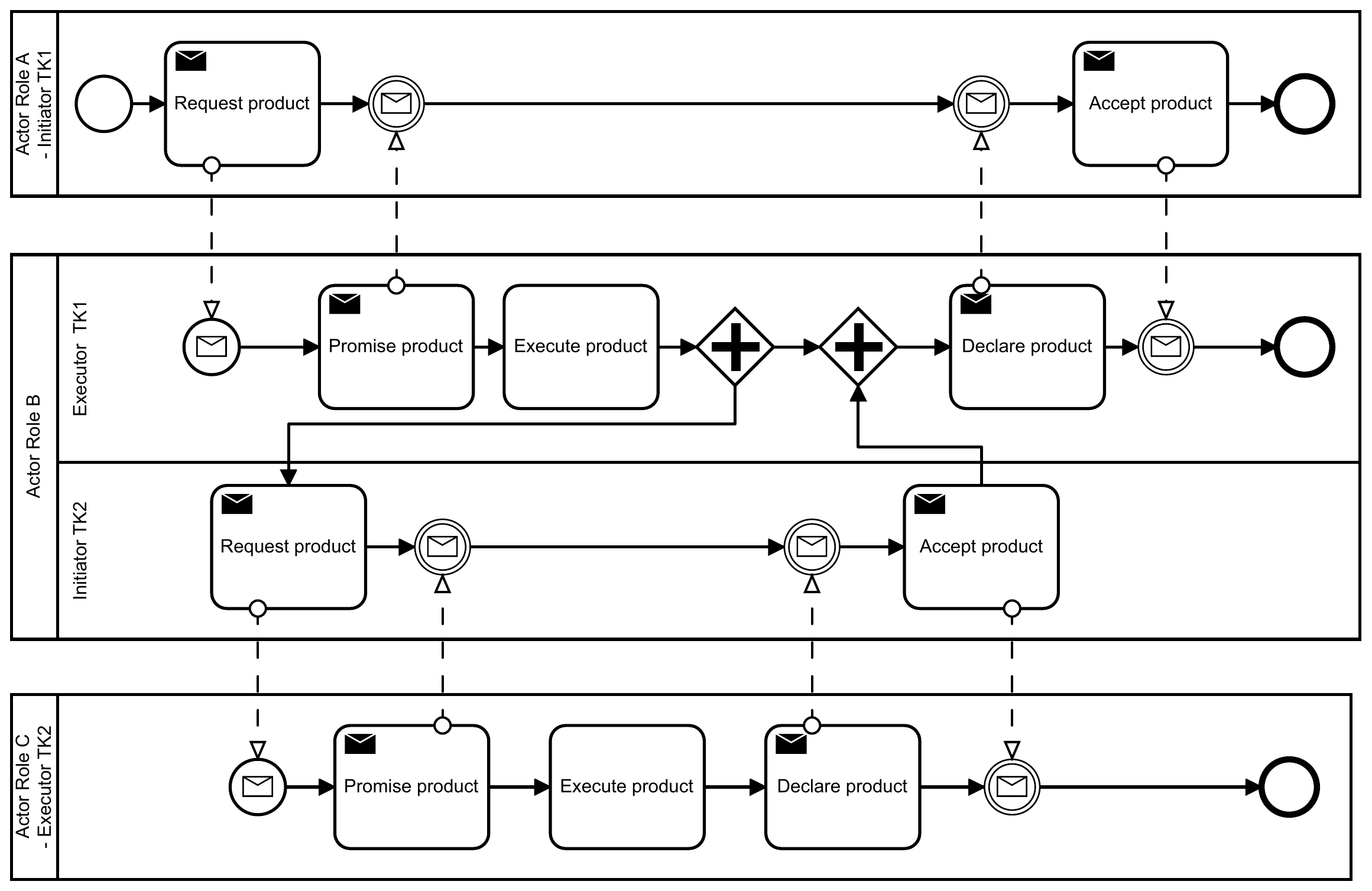}	}
      \caption{Dependency between transactions: request after execution (RaE) composition pattern concerning 2 business transactions.}
        \label{demo.patternRaE}
\end{figure}

\begin{figure}[htbp]
		\centering { \includegraphics[width=0.45\textwidth]{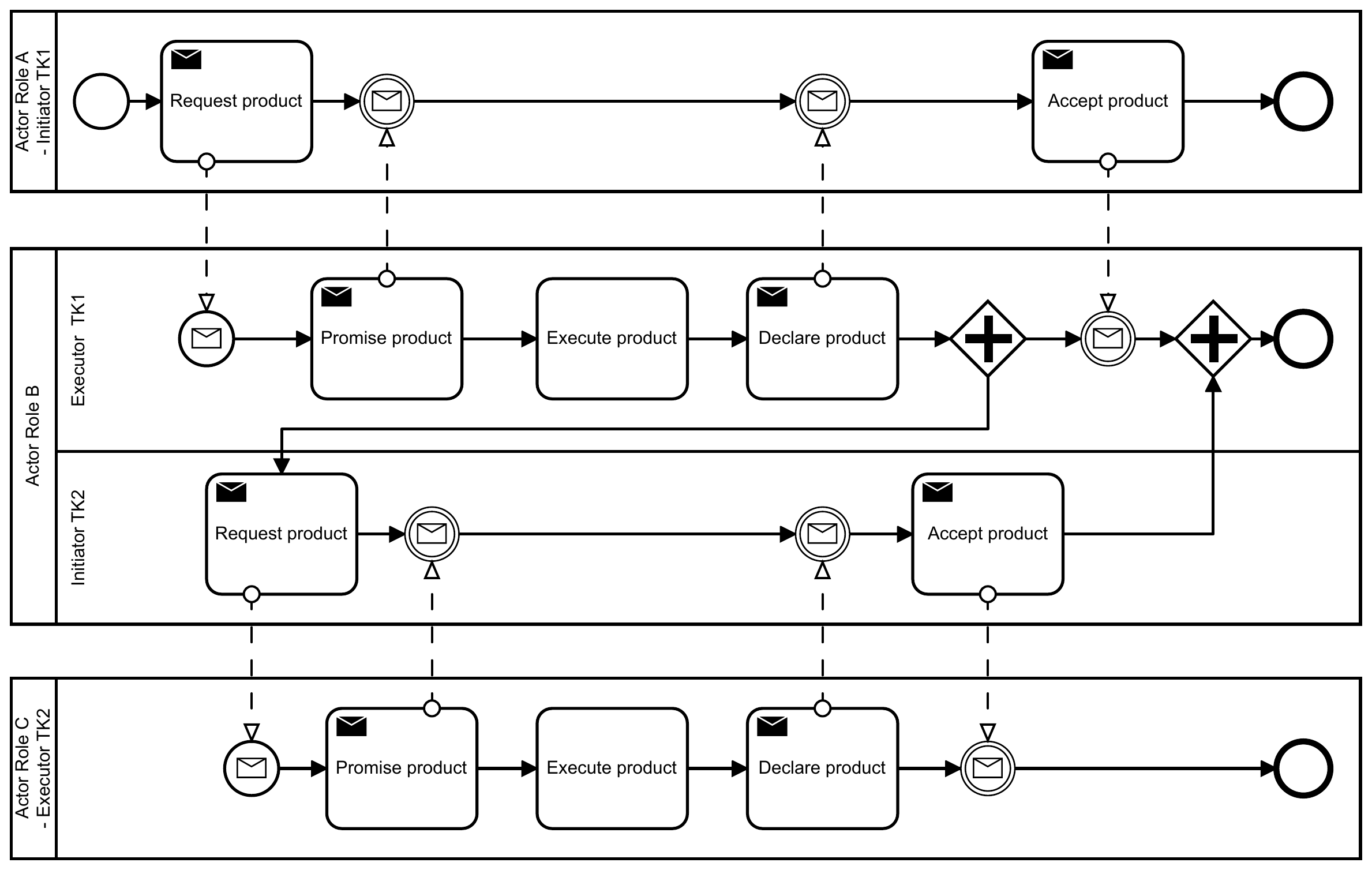}	}
      \caption{Dependency between transactions: request after declare (RaD) composition pattern concerning 2 business transactions.}
        \label{demo.patternRaD}
\end{figure}

A composition pattern occurs whenever an executor actor role assumes the responsibility of initiating another transaction.
Figure~\ref{demo.patternRaP2tx} exemplifies a dependency between the \emph{``happy flow''} of two business transactions using the request after promise (RaP) pattern. It is noted in this example the actor role B assumes the role of Transaction 1 TE and Transaction 2 TI. The coordination between them is performed by the correspondingly BPMN process. For model simplification, the declination, the rejection and the revocations are not designed in the model. However, the dependency is exactly the same.
Figure~\ref{demo.patternRaP3tx} depicts three transactions using the RaP as dependency pattern. Transaction 1 is declared when Transaction 2 is accepted, and consequently, Transaction 2 is declared only after Transaction 3 has been accepted. This pattern is usual used when the production responsibility need to be assigned to other actor role, \emph{e.g.}, outsourcing.
Figure~\ref{demo.patternRaP3txsimul} shows a multiple, and concurrent, invocation of transactions.
After Transaction 1 had been promised, then Transaction 2 and Transaction 3 are requested. Conversely, Transaction 1 is only declared after the declaration of Transaction 2 and Transaction 3. In this example, it is noted that an actor is assigned with three roles, \emph{e.g.}, Actor Role B is Transaction 1 TE, Transaction 2 TI and Transaction 3 TI.
In Figure~\ref{demo.patternRaE} is depicted a pattern where the second transaction is started only after the execution of the first: the request after execution (RaE). This pattern is useful whenever a synchronous sequence between transactions is required. When asynchronous sequence is expected the pattern request after declare (RaD) could be used (\emph{cf.} exemplified in Figure~\ref{demo.patternRaD}).     

Summarizing, DEMO $\Psi$-theory specifies the principles to assign semantic meaning to the structure and dynamic of business transactions, encompassing the actors, the communications, the productions and all its dependencies. On the contrary, BPMN does not prescribe any semantic for the business process model; it only provides a set of constructs that could be combined accordingly with a specification.  
The following core concerns were taken in consideration to map the DEMO complete pattern onto BPMN:
\emph{(i)} each actor role is autonomous, therefore, a BPMN pool per each one is required,
\emph{(ii)} all the communication acts between actors are fully specified considering the supporting gateways and events that are required, and
\emph{(iii)} the DEMO business transaction pattern is repeated recurrently where the patterns are key to integrate complex business transactions networks.

From a practical perspective, the required inputs to semantify business processes are: \emph{(i)} a list of actors, \emph{(ii)} a list of transactions where the actors are involved and specifying the result expected, and \emph{(iii)} the dependencies between the transactions, namely, RaP, RaE or RaD.
An open-source software prototype is publicly available online\footnote{\url{https://github.com/SemantifyingBPMN/SemantifyingBPMN}}.  To facilitate the rendering, each business transaction view, can be configured as happy flow or happy flow + declinations and rejections or complete pattern.

\section{Proof-of-concepts and discussion}
\label{omegaA.validation}

This section applies the mapping of DEMO complete pattern onto BPMN models on two different networks of business processes and discusses the achieved results. 
Both proof-of-concepts (PoC) are based on documentation that reflects the discussion and understanding between the stakeholders referring to \emph{a)} a process of payment approval and \emph{b)} a process of budgetary change request. However, both are poorly documented in BPMN, containing syntax and semantic errors, and different assumptions exists between them. 
These models were implemented on a BPM engine with a specification far beyond from their initial specification, leaving an untraceable link between the process model and its implementation. This has an immediate impact on testing and acceptance of current process implementation, but also on future changes and/or adaptations.
In this context, our proposal reveals many hidden or implicit steps in the original BPMN model, mitigating the gaps between the model and its implementation. A key issue in the the proof-of-concepts, relies on the fact that the process engine has a couple of pre-existing features that an actor can do on every task of the process instance: 
\emph{(i)} cancel current task and redo previous task, and 
\emph{(ii)} stop the process (abort).
Such options are not in the original BPMN model, but are used in practice to support, \emph{e.g.}, decline decisions. Since they do not require any new code to be developed, because it is supported natively by the process engine (unless explicitly disabled), they are implicitly assumed. However, such implicit decline possibilities has an major impact on process model, testing, and overall present and future understanding.  
 
The goal of this section is to show the usefulness of our proposal, measuring the coverage, both positive and negative, of the mapping of DEMO complete pattern onto the PoCs' BPMN models.

\subsection{Methodology for proof-of-concepts}
The following steps were applied to both PoCs. 

\begin{enumerate}
\item Firstly, the available documentation from each PoC was reviewed. Each one has a requirement specification document resulting from the interactions with each customer (named as \emph{(doc.a)}).
In addition, a \emph{(doc.b)} containing a simplified BPMN model designed from \emph{(doc.a)} was reviewed.

\item Afterwards, \emph{(doc.c)} that is a spreadsheet with business objects containing details about some BPMN tasks and further technological integration details was reviewed. Due to the introduction of unnecessary complexity for our paper, \emph{(doc.c)} is not considered.  

\item Then, the \textbf{actor roles} are identified, using \emph{(doc.a)} and \emph{(doc.b)}.
After that, the \textbf{DEMO business transactions} and correspondingly expected \textbf{results} are identified. For each transaction the actor role was classified as either \textbf{initiator} or \textbf{executor}. 
Then, the \textbf{dependencies} between transactions (namely RaP, RaD or RaE) are identified.

\item From previous step and using the proposal explained in Section~\ref{omegaA.foundationaltheory}, a semantified BPMN model is produced.

\item After that, a matrix identifying the explicit DEMO pattern coverage is produced as an outcome of the PoC. The \emph{(doc.b)} specifies an incomplete \emph{``happy flow''}, therefore only a partial coverage is obtained in this step in both PoCs. 

\item To enrich the previous matrix with the implicit DEMO pattern coverage, we used interviews with the BPM engine experts. Implicit corresponds to the transaction steps and communication acts that were not modeled, but, are assumed as existing during BPM engine implementation. This duality between explicit and implicit parts of a model has already been referred in the literature~\cite{caetano2012using}~\cite{pascoa2011ontology}. 	 
\end{enumerate}

\subsection{Proof-of-concept 1 description}

%\input{sections/usecaseAExplicit}
%\paragraph{Semantifying use case A without enrichments considering the explicit, and implicit, communication and production acts}

The context of PoC 1 is collected from a public institution and consists in one business process with the goal of operating a budget transfer between business units. For that end, the SOC Department business unit solicits the budget transfer need, and then, a four-level validation chain is executed by the following sequence of business units:  
SPFP,
SPFP Coordinator,
Financial Department Director, and
Governing Board Representative.		
Each business unit is represented as an actor role and are described in Table~\ref{usecaseA.actors}.

\begin{table*}[htpb]
	\centering
%		\resizebox{\textwidth}{!}{%
			\scalebox{1}{
			\small
	\begin{tabular}{p{6cm}|p{11cm}}
	\textbf{Actor}	& \textbf{Description}   \\\hline
		\textbf{A01} - SOC Department & Budget transfer solicitor.   \\\hline
		\textbf{A02} - SPFP					&  Executor of the first validation and creator of a budget transfer proposal.\\\hline
		\textbf{A03} - SPFP Coordinator		&  Executor of the second validation of the budget transfer.  \\\hline
		\textbf{A04} - Financial Department Director		&   Executor of the third validation of the budget transfer. \\\hline
		\textbf{A05} - Governing Board Representative				&   Executor of the fourth validation of the budget transfer.  \\\hline
	\end{tabular}
}
\caption{Actor roles identification, proof-of-concept 1}
\label{usecaseA.actors}
\end{table*}

When the SOC Department initiates the solicitation of a budget change, SPFP is the actor role responsible to execute this transaction (TK01). In the end of TK01, a new transaction result is expected in the world: the \emph{[budget] that has been changed}. Meaning that the product kind \emph{budget} has been changed accordingly with the initial solicitation.
Since SPFP depends on the validation of the SPFP Coordinator, then TK02 is initiated. The expected result of this new transaction, provided by the SPFP Coordinator, corresponds to the \emph{[budget change] has been validated by second level}.
Afterwards, TK03 and TK04 are initiated sequentially by SPFP Coordinator and Financial Department Director, respectively. The execution' responsibility is assigned, correspondingly, to the actor roles of Financial Department Director and Governing Board Representative.
The expected transaction results are the \emph{[budget change] has been validated by third level}, and the \emph{[budget change] has been validated by fourth level}.
This set of transaction kinds are presented in Table~\ref{usecaseA.transactions}.

\begin{table*}[htpb]
		\centering
				\resizebox{\textwidth}{!}{%
	\begin{tabular}{p{6cm}| p{5cm} |p{5cm} |p{5cm}}
		\textbf{Transaction kind}	& \textbf{Initiator role} & \textbf{Executor role} & \textbf{Transaction result with [product kind]} \\\hline
		\textbf{TK01} - Soliciting budget change  &  A01 - SOC Department 		&	A02 - SPFP			& \textbf{PK01} [budget] has been changed	  \\\hline
		
		\textbf{TK02} - Validating second-level budget change	&  	A02 - SPFP	&	A03 - SPFP Coordinator	& \textbf{PK02} [budget change] has been validated by second level  \\\hline
		
		\textbf{TK03} - Validating third-level budget change &  	A03 - SPFP Coordinator		&	A04 - Financial Department Director	& \textbf{PK03} [budget change] has been validated by third level  \\\hline
		
		\textbf{TK04} - Validating fourth-level budget change	&  	A04 - Financial Department Director	&	A05 - Governing Board Representative	& \textbf{PK04} [budget change] has been validated by fourth level  \\\hline
	\end{tabular}
	}
	\caption{Transaction kinds identification, proof-of-concept 1}
	\label{usecaseA.transactions}
\end{table*}

Then, Table~\ref{usecaseA.dependencies} specifies the dependencies between business transaction. From \emph{(doc.a)} and \emph{(doc.b)} we obtained that TK02 is initiated by TK01 after its promise (RaP), and TK03 is initiated by TK02 after validation execution (RaE), and the same applies to TK04 that is initiated by TK03 after validation execution (RaE).
Those dependencies defines how should the business transactions be connected in a chain to operate the desired business process.

\begin{table}[htpb]
		\centering
			\resizebox{0.3\textwidth}{!}{%
	\begin{tabular}{|p{2cm}|c|c|c|c|}
	\hline
		& \multicolumn{4}{c|}{\emph{Calling transaction kind}}\\
	\emph{Invoked transaction kind}
					& \textbf{TK01} & \textbf{TK02} & \textbf{TK03} & \textbf{TK04} \\\hline
	\textbf{TK01}	&  				&				&				&				  \\\hline
	\textbf{TK02}	&  		RaP		&				&				&				  \\\hline
	\textbf{TK03}	&  				&		RaE		&				&				 \\\hline
	\textbf{TK04}	&  				&				&	RaE			&				  \\\hline
	\end{tabular}
}
	\caption{Transaction kinds dependencies identification, proof-of-concept 1. Where, RaP = Request after Promise pattern, RaE = Request after Execution, RaD = Request after Declare pattern.}
	\label{usecaseA.dependencies}
\end{table}

\begin{figure}[tbhp]
	\centering
	\includegraphics[width=0.5\textwidth]{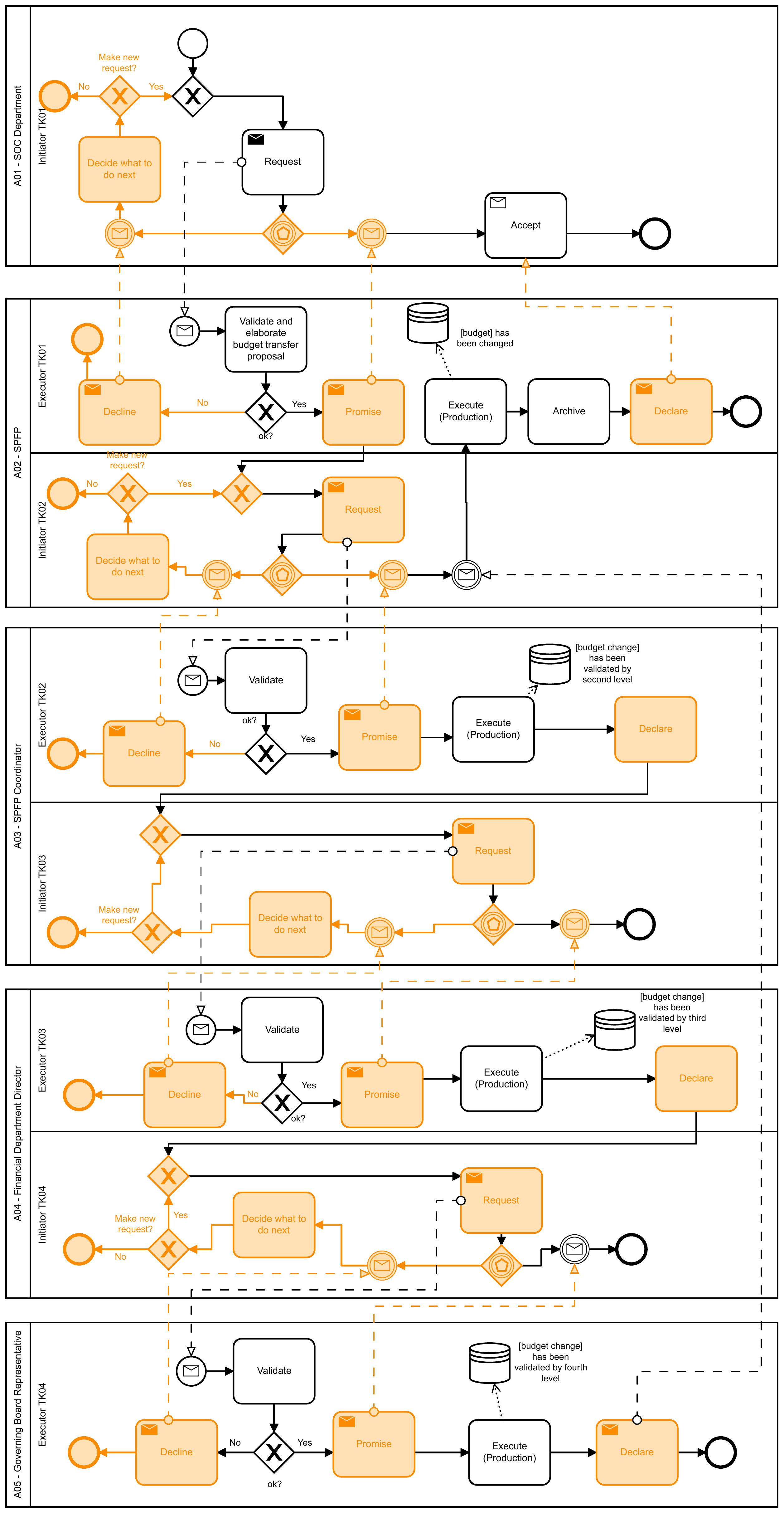}
	\caption{Proof-of-concept 1 semantified with explicit and implicit identification of elements. Explicit elements are depicted in black, and implicit are in \textcolor{orange}{orange}.}
	\label{usecaseA.semantified.before}
\end{figure}

This PoC' network of business transactions is then designed targeting the full usage of the DEMO complete pattern.
However, few details exists explicitly in \emph{(doc.a)} and \emph{(doc.b)} specification, to design an informed semantified BPMN. The explicit definitions are represented in black color in Figure~\ref{usecaseA.semantified.before}, and as it could be noticed only the executions, validations, one archive, one initial request and one accept are depicted.
Therefore, using the semantifying BPMN tool\footnote{The provided software tool only requires the data input available from Tables~\ref{usecaseA.actors}~\ref{usecaseA.transactions}~\ref{usecaseA.dependencies}}, we enrich the PoC 1 model with the semantified BPMN pattern to fill the missing gaps.
With this solution, a more complete model is depicted in the overall of Figure~\ref{usecaseA.semantified.before}.  
Yet, for presentation simplification, the declinations and rejections are exemplified, but the revocations are not.

\subsection{Proof-of-concept 2 description}

%\input{sections/usecaseBExplicit}
%\paragraph{Semantifying use case B without enrichments considering the explicit, and implicit, communication and production acts}

The proof-of-concept 2 consists in the business process of payment authorization and execution. It is operated in the same public institution as introduced in PoC 1. 
The following seven business units are involved:
Financial Department Registrar, 
Financial Department Expense Control, 
Financial Department Coordinator, 
Financial Department Director, 
Governing Board Representative, 
Governing Board, and 
Financial Department treasury.
Each business unit is represented as an actor role and are described in Table~\ref{usecaseB.actors}.

\begin{table*}[htpb]
	\centering
	%\resizebox{\textwidth}{!}{%
	\scalebox{1}{
	\small
	\begin{tabular}{p{6cm}|p{11cm}}
	\textbf{Actor}	& \textbf{Description}   \\\hline
		\textbf{A01} - Financial Department Registrar 	& Autorization payment solicitator.   \\\hline
		\textbf{A02} - Financial Department Expense Control	& Executes the public administration payment requirements. \\\hline
		\textbf{A03} - Financial Department Coordinator		& Validate the public administration payment requirements and solicits the payment authorization.\\\hline
		\textbf{A04} - Financial Department Director		& Payment authorization for amount $< 5k\euro$. \\\hline
		\textbf{A05} - Governing Board Representative 				& Payment authorization for amount between $>= 5k\euro$ and $< 30k\euro$.		    \\\hline
		\textbf{A06} - Governing Board						& Payment authorization for amount $>= 30k\euro$.   \\\hline
		\textbf{A07} - Financial Department Treasury & Executes previous checked and authorized payments. \\\hline
	\end{tabular}
}
\caption{Actor roles identification, proof-of-concept 2.}
\label{usecaseB.actors}
\end{table*}

The Financial Department Registrar starts by soliciting a payment authorization (TK01), which is executed by the Financial Department Expense Control, delivering the result: \emph{the [authorization] has been satisfied}. 
To fulfill this result, the Financial Department Expense Control triggers TK02 with a dependency RaP (of TK01) and, later, will initiate TK06 with dependency RaE (of TK01).
Financial Department Coordinator is the responsible for TK02 execution and the expected transaction result is that \emph{[payment requirement] has been validated}. If this result is achieved, then TK03 is initiated with a RaE dependency. 
TK03 verifies the amount to be authorized. If it is below $5k\euro$ then it could be authorized, otherwise TK04 or TK05 will be initiated accordingly with the amount between $>=5k\euro$ and $<30k\euro$, or $>=30k\euro$.
TK03, TK04 and TK05 are all related with the \emph{[payment] authorization} result, but differ in the amount considered.
When one of the TK03, TK04 or TK05 are completed, then the control is back to TK03 and afterwards to TK01 that initiates TK06 to reach the desired result of \emph{[payment] has been executed}.

This set of transaction kinds are presented in Table~\ref{usecaseB.transactions}, and the dependencies between transactions are in Table~\ref{usecaseB.dependencies}.

\begin{table*}[htpb]
		\centering
		\resizebox{\textwidth}{!}{%
	\begin{tabular}{p{6cm}| p{5cm} |p{5cm} |p{5cm}}
		\textbf{Transaction kind}	& \textbf{Initiator role} & \textbf{Executor role} & \textbf{Transaction result with [product kind]} \\\hline	
		
		\textbf{TK01} - Soliciting payment authorization	& 	A01 - Financial Department Registrar	&	A02 - Financial Department Expense Control		&	\textbf{PK01} - the [authorization] has been satisfied	\\\hline 
	
		\textbf{TK02} - Executing the public administration payment requirements	& A02 - Financial Department Expense Control & A03 - Financial Department Coordinator			& \textbf{PK02} - the [payment requirement] has been validated  \\\hline 
		
		\textbf{TK03} - Authorizing the payment $<5k\euro$ 			&  	A03 - Financial Department Coordinator		& A04 - Financial Department Director		& \textbf{PK03} - the [payment] $<5k\euro$ has been authorized  \\\hline 
		
		\textbf{TK04} - Authorizing the payment $>5k\euro$ and $<30k\euro$ & A04 - Financial Department Director 		& A05 - Governing Board Representative  		& \textbf{PK04} - the [payment] $>=5k\euro$ e $<30k\euro$ has been authorized  \\\hline 
		
		\textbf{TK05} - Authorizing the payment $>30k\euro$ 			& A04 - Financial Department Director		& A06 - Governing Board				&  \textbf{PK05} - the [payment] $>=30k\euro$ has been authorized  \\\hline  
		
		\textbf{TK06} - Paying 				&	A02 - Financial Department Expense Control	& A07 - Financial Department Treasury		& \textbf{PK06} - the [payment] has been executed\\\hline
	\end{tabular}
	}
	\caption{Transaction kinds identification, proof-of-concept 2.}
	\label{usecaseB.transactions}
\end{table*}

\begin{table}[htpb]
		\centering
			%\resizebox{\textwidth}{!}{%
	\scalebox{0.7}{
	\small

	\begin{tabular}{|p{2cm}|c|c|c|c|c|c|}
	\hline
		& \multicolumn{6}{c|}{\emph{Calling transaction kind}}\\
	\emph{Invoked transaction kind}
					& \textbf{TK01} & \textbf{TK02} & \textbf{TK03} & \textbf{TK04} & \textbf{TK05} & \textbf{TK06} \\\hline
	\textbf{TK01}	&  				&				&				&				&				&				\\\hline
	\textbf{TK02}	&  	RaP			&				&				&				&				&				\\\hline
	\textbf{TK03}	&  				&		RaE		&				&				&				&				\\\hline
	\textbf{TK04}	&  				&				&		RaP		&				&				&				\\\hline
	\textbf{TK05}	&  				&				&		RaP		&				&				&				\\\hline
	\textbf{TK06}	&  	RaE			&				&				&				&				&				\\\hline
	\end{tabular}
}
	\caption{Transaction kinds dependencies identification, proof-of-concept 2. Where, RaP = Request after Promise pattern, RaE = Request after Execution, RaD = Request after Declare pattern. }
	\label{usecaseB.dependencies}
\end{table}

\begin{figure*}[htpb]
	\centering
	\includegraphics[width=0.6\textwidth]{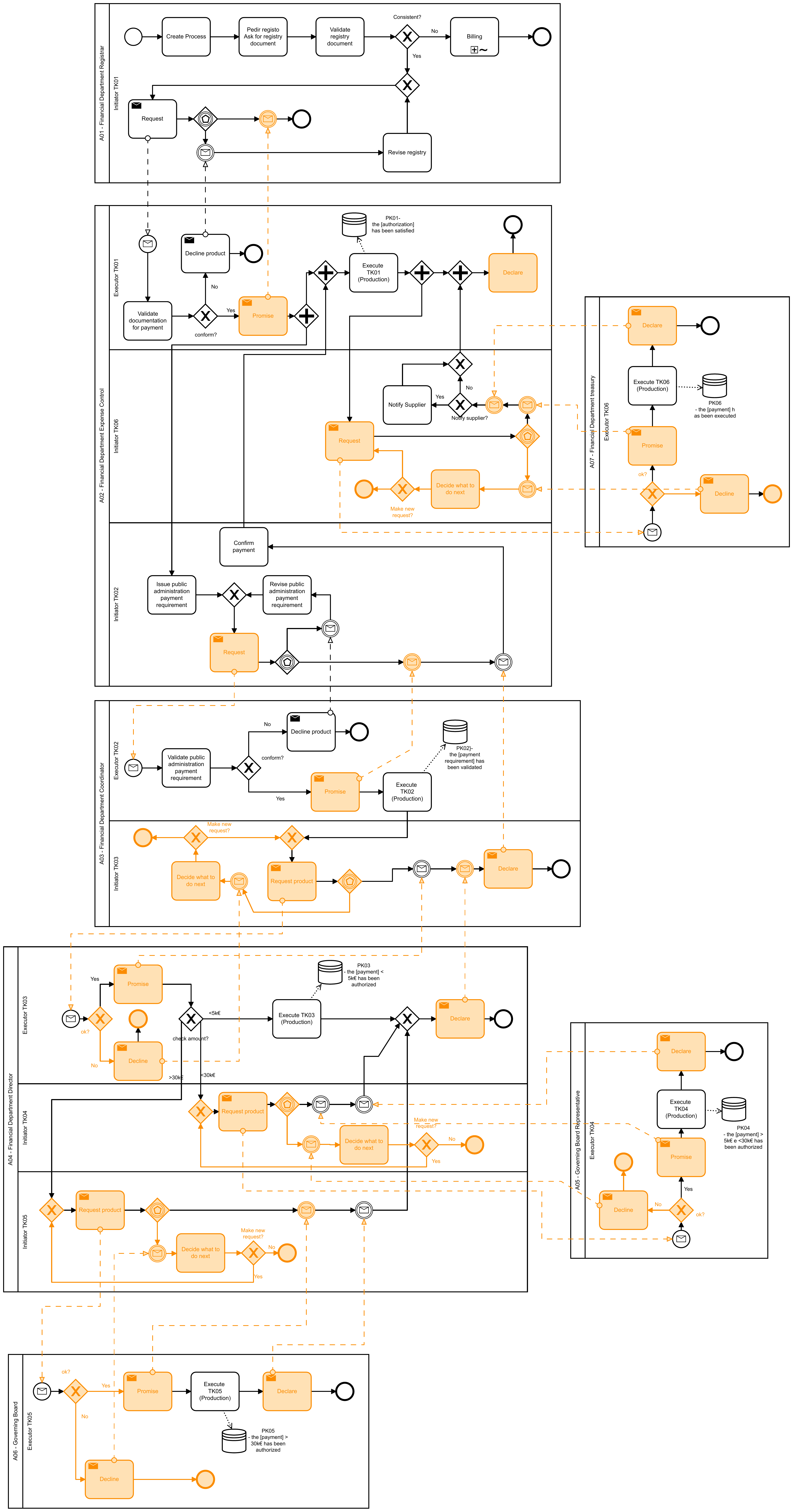}
	\caption{Proof-of-concept 2 semantified with explicit and implicit identification of elements. Explicit elements are depicted in black, and implicit are in \textcolor{orange}{orange}.}
	\label{usecaseB.semantified.before}
\end{figure*}

As explained in PoC 1, the PoC2 network of business transactions is designed using the explicit details available in \emph{(doc.a)} and \emph{(doc.b)} and represented in black color in Figure~\ref{usecaseB.semantified.before}. Again, only a small amount of elements are depicted. 
Enriching this model with the semantified BPMN pattern (including declinations and rejections) offered the representation of the orange elements in the same Figure.

\subsection{Results discussion}

Both BPMN models (\emph{cf.} Figures~\ref{usecaseA.semantified.before} and~\ref{usecaseB.semantified.before}) were presented, and explained, to the BPM engine experts during one interview and by email. 
Then, the experts were asked to identify which acts are implemented, and which are not, within the BPM engine.
In specific, we inquired if the semantifed BPMN complete pattern is implemented or if some parts could be considered as implicit. 
Otherwise, it was considered as not implemented.
The result of this assesment is summarized in Tables~\ref{usecaseA.checkcoverage} and~\ref{usecaseB.checkcoverage}. They display a matrix with all transaction kinds related with all the possible production and communication acts. 
Black dots represent the acts that were explicitly modeled since the initial BPMN specified in \emph{(doc.b)}. 
In orange dots are the acts that were found implemented but not modeled, and an empty dot the acts that are neither implemented nor modeled.

\begin{table}[htpb]
	\centering
	\resizebox{0.5\textwidth}{!}{%
		\begin{tabular}{|p{3cm}|c|c|c|c|c|c|}
			\hline
			& \multicolumn{4}{c|}{\emph{Transaction kinds}} && \\	
			\emph{Communication and Production acts} & 
			\textbf{TK01} 	& \textbf{TK02} 	& \textbf{TK03} 		& \textbf{TK04} 	&& \begin{tabular}[c]{@{}l@{}}Sum\\ (\CircleSolid/\textcolor{orange} {\CircleSolid}/\CircleShadow)\end{tabular}\\\hline
			Request	  	&	\CircleSolid	&	\textcolor{orange} {\CircleSolid}	&	\textcolor{orange} {\CircleSolid}		&	\textcolor{orange} {\CircleSolid} 	&&	(1/3/0)	\\\hline
			Promise		&  	\textcolor{orange}	{\CircleSolid}			&	\textcolor{orange}	{\CircleSolid}				&	\textcolor{orange}	{\CircleSolid}					&	\textcolor{orange}	{\CircleSolid}	 			&&	(0/4/0)	\\\hline
			Execute  	&	\CircleSolid	&	\CircleSolid	&	\CircleSolid		&  	 \CircleSolid	&&	(4/0/0)	\\\hline
			Declare 	&  	\textcolor{orange}	{\CircleSolid}	&	\textcolor{orange}	{\CircleSolid}	&	\textcolor{orange}	{\CircleSolid}	&	\textcolor{orange}	{\CircleSolid} 	&&	(0/4/0)	 \\\hline
			Accept 		&  	\CircleSolid	&	\CircleShadow	&	\CircleShadow		&	\CircleShadow	&&	(1/0/3) \\\hline
			Decline 	&  \textcolor{orange}	{\CircleSolid}	&\textcolor{orange}	{\CircleSolid}	&\textcolor{orange}	{\CircleSolid}	&\textcolor{orange}	{\CircleSolid}	&&	(0/4/0) \\\hline
			Reject 		&  		\textcolor{orange}	{\CircleSolid}			&	\textcolor{orange}	{\CircleSolid}				&	\textcolor{orange}	{\CircleSolid}					&	\textcolor{orange}	{\CircleSolid}	 			&&	(0/4/0) \\\hline
			Revoke Request&  	\CircleShadow			& \CircleShadow						&	\CircleShadow						&	\CircleShadow		 			&&	(0/0/4) \\\hline
			Revoke Promise&  \CircleShadow					&	\CircleShadow					&	\CircleShadow						&	\CircleShadow		 			&&	(0/0/4) \\\hline
			Revoke Declare&  	\CircleShadow				&	\CircleShadow					&	\CircleShadow						&	\CircleShadow		 			&&	(0/0/4) \\\hline
			Revoke Accept&  \CircleShadow					&	\CircleShadow					&\CircleShadow							&\CircleShadow			 			&&	(0/0/4) \\\hline
			Allow	  	&  	\CircleShadow					&	\CircleShadow					&	\CircleShadow						&	\CircleShadow		 			&&	(0/0/4) \\\hline
			Stop		&  	\CircleShadow					&	\CircleShadow					&	\CircleShadow						&	\CircleShadow		 			&&	(0/0/4) \\\hline
			Refuse 		&  	\CircleShadow					&	\CircleShadow					&	\CircleShadow						&\CircleShadow			 			&&	(0/0/4) \\\hline\hline
			Sum (\CircleSolid/\textcolor{orange} {\CircleSolid}/\CircleShadow)	&		(3/4/7)	&		(1/5/8)	&			(1/5/8)	& 	(1/5/8)		&&			\\\hline
			\multicolumn{7}{|r|}{Total Implemented = 25 (in 56) = 44,6\%, where}\\
			\multicolumn{7}{|r|}{Total \textbf{Explicit} = 6 (in 56) = 10,7\%}\\
			\multicolumn{7}{|r|}{Total \textcolor{orange} {Implicit} = 19 (in 56) = 33,9\%}\\\hline
		\end{tabular}
	}
	\caption{Semantified BPMN using complete DEMO pattern with explicit and implicit acts, proof-of-concept 1. Where, \CircleSolid~represents an explicit act, \textcolor{orange} {\CircleSolid}~represents an implicit act, \CircleShadow~represents a not implemented act.}
	\label{usecaseA.checkcoverage}
\end{table}

\begin{table*}[htpb]
	\centering
	\resizebox{0.7\textwidth}{!}{%
		
		\begin{tabular}{|p{3cm}|c|c|c|c|c|c|p{0.01cm}|c|}
			\hline
			& \multicolumn{6}{c|}{\emph{Transaction kinds}} && \\	
			\emph{Communication and Production acts} & 
			\textbf{TK01} 	& \textbf{TK02} 	& \textbf{TK03} 		& \textbf{TK04} 	& \textbf{TK05} & \textbf{TK06} && \begin{tabular}[c]{@{}l@{}}Sum\\ (\CircleSolid/\textcolor{orange} {\CircleSolid}/\CircleShadow)\end{tabular}\\\hline
			Request	  	&	\CircleSolid	&	\textcolor{orange} {\CircleSolid}	&\textcolor{orange} {\CircleSolid}	&	\textcolor{orange} {\CircleSolid} &\textcolor{orange} {\CircleSolid}&	\textcolor{orange} {\CircleSolid}&&	(1/5/0)	\\\hline
			Promise		&  	\textcolor{orange} {\CircleSolid}	&	\textcolor{orange} {\CircleSolid}	&\textcolor{orange} {\CircleSolid}	&		\textcolor{orange} {\CircleSolid}	&\textcolor{orange} {\CircleSolid}	&	\textcolor{orange} {\CircleSolid}			&&	(0/6/0)	\\\hline
			Execute  	&	\CircleSolid	&	\CircleSolid	&	\CircleSolid		& \CircleSolid	 	&\CircleSolid	&	\CircleSolid&&	(6/0/0)	\\\hline
			Declare 	&  	\textcolor{orange} {\CircleSolid}				&	\textcolor{orange} {\CircleSolid}				&	\textcolor{orange} {\CircleSolid}		&\textcolor{orange} {\CircleSolid}	&\textcolor{orange} {\CircleSolid}	&\textcolor{orange} {\CircleSolid}	&&	(0/6/0)	 \\\hline
			Accept 		&  \CircleShadow	&	\CircleShadow	&	\CircleShadow		&	\CircleShadow	&\CircleShadow	&\CircleShadow	&&	(0/0/6) \\\hline
			Decline 	&  \CircleSolid		&	\CircleSolid	&	\textcolor{orange} {\CircleSolid}		&	\textcolor{orange} {\CircleSolid}	 			&	\textcolor{orange} {\CircleSolid}			&	\textcolor{orange} {\CircleSolid}			&&	(2/4/0) \\\hline
			Reject 		&  		\textcolor{orange} {\CircleSolid}				&	\textcolor{orange} {\CircleSolid}					&	\textcolor{orange} {\CircleSolid}						&	\textcolor{orange} {\CircleSolid}		 			&		\textcolor{orange} {\CircleSolid}			&	\textcolor{orange} {\CircleSolid}				&&	(0/6/0) \\\hline
			Revoke Request&  	\CircleShadow				&	\CircleShadow					&	\CircleShadow						&	\CircleShadow		 			&	\CircleShadow				&	\CircleShadow				&&	(0/0/6) \\\hline
			Revoke Promise&  				\CircleShadow				&	\CircleShadow					&	\CircleShadow						&	\CircleShadow		 			&	\CircleShadow				&	\CircleShadow				&&	(0/0/6) \\\hline
			Revoke Declare&  					\CircleShadow				&	\CircleShadow					&	\CircleShadow						&	\CircleShadow		 			&	\CircleShadow				&	\CircleShadow				&&	(0/0/6) \\\hline
			Revoke Accept&  				\CircleShadow				&	\CircleShadow					&	\CircleShadow						&	\CircleShadow		 			&	\CircleShadow				&	\CircleShadow				&&	(0/0/6) \\\hline
			Allow	  	&  					\CircleShadow				&	\CircleShadow					&	\CircleShadow						&	\CircleShadow		 			&	\CircleShadow				&	\CircleShadow				&&	(0/0/6) \\\hline
			Stop		&  					\CircleShadow				&	\CircleShadow					&	\CircleShadow						&	\CircleShadow		 			&	\CircleShadow				&	\CircleShadow				&&	(0/0/6) \\\hline
			Refuse 		&  					\CircleShadow				&	\CircleShadow					&	\CircleShadow						&	\CircleShadow		 			&	\CircleShadow				&	\CircleShadow				&&	(0/0/6) \\\hline\hline
			Sum (\CircleSolid/\textcolor{orange} {\CircleSolid}/\CircleShadow)	&		(3/3/8)	&		(2/4/8)	&			(1/5/8)	& 	(1/5/8)		&	(1/5/8)	&	(1/5/8)	&&	 		 \\\hline
			\multicolumn{9}{|r|}{Total Implemented = 24 (in 56) = 42,9\%, where}\\
			\multicolumn{9}{|r|}{Total \textbf{Explicit} = 7 (in 56) = 12,5\%}\\
			\multicolumn{9}{|r|}{Total \textcolor{orange} {Implicit} = 17 (in 56) = 30,4\%}\\\hline
		\end{tabular}
	}
	\caption{Semantified BPMN using complete DEMO pattern with explicit and implicit acts, proof-of-concept 2. Where, \CircleSolid~represents an explicit act, \textcolor{orange} {\CircleSolid}~represents an implicit act, \CircleShadow~represents a not implemented act.}
	\label{usecaseB.checkcoverage}
\end{table*}

Therefore, relating the produced semantified BPMN model with the two sources of \emph{(doc.b)} and experts knowledge, allows the completeness asessment of the existing models and its correspondingly implementation.
Moreover, it could be used to identify which parts of the model is not covered, and then generate it automatically and afterwards to establish an enrichment plan to the BPM engine.

Finally, it has been identified that the BPM engine implementation can only support revocations when the actor role owns the execution of the act itself. Therefore, we consider that this is too restricting to enforce the full revocation patterns, and thus, it is considered as not implemented.
To summarize, in the context of this specific BPM engine, we pose the following two questions.
\paragraph{What are the communication and production acts that are explicitly modeled and at the same time implemented in the BPM engine?}
Execution is the consensual act that is modeled and implemented.
Also, an initial request, that triggers the business process, could be found in both PoCs and implemented.
Then, depending in the PoC, an accept and a decline is found in the model and also explicitly implemented. Experts were queried about this and it was found that different teams of modelers and developers were involved in the projects, therefore different criteria was applied. This paper advocates that the usage of a semantified pattern can facilitate the normalization of criteria between projects involving different teams.
The accept is a communication act that is considered only once in PoC2. Accordingly with the experts, this was a requirement not sought often by customers.
\paragraph{What coverage of the semantified BPMN pattern has been identified implicitly in the PoCs?}
In terms of implicit acts many exists in the BPM engine (request, promise, declare, decline and reject) what could be considered as an oversimplification of the BPMN models that could bias the project' understanding and the communication with customers.
For instance, since revocation are not implemented, they could be confused with the remaining implicit acts.
Under the lens of a explicit business process model, it is not possible to distinguish between an implicit act or a not implemented act.

\section{Related work}
\label{omegaA.relatedwork}

\begin{figure}[htbp]
		\centering { \includegraphics[width=0.4\textwidth]{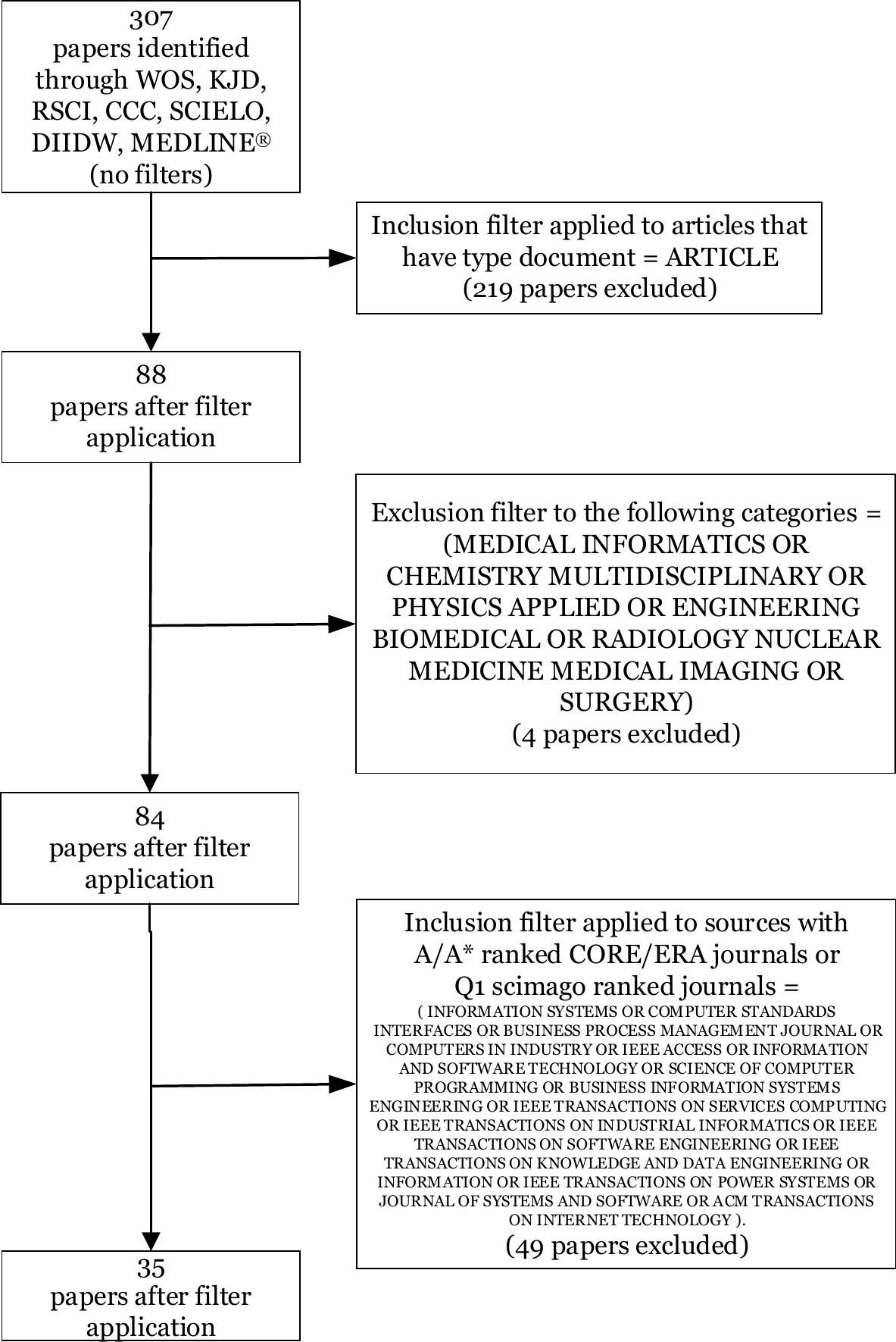}	}
       \caption{Filtration process for the related work.}
        \label{filters}
\end{figure}

This section assesses the innovation of our proposal performing a systematic review of the knowledge available in the literature related to the topics of BPMN patterns and BPMN semantic.
To that end, a search has been conducted using the  Web of Science Core Collection (WOS), KCI-Korean Journal Database (KJD), Russian Science Citation Index (RSCI), Current Contents Connect (CCC), SciELO Citation Index (SCIELO), and MEDLINE\textsuperscript{\textregistered} databases, considering the following topic search: TS = (\emph{``BPMN  AND (semantic  OR pattern)''}), until 2020 (included). 
Due to the large number of hits returned, the search has been filtered as depicted in Figure~\ref{filters}: 
including only articles;
excluding non-related categories and 
including only A-ranked CORE/ERA or Q1 scimago ranked journals. 
For easier reference, the full list of bibliographic references obtained with the search is presented in Table~\ref{list.publications} and the distribution by year is presented in Table~\ref{table.mountbyyear}.
The quality of this retrieved TS is given by the total number of citations (excluding auto-citations) of 664, the average citations \emph{per} item of 19.26 and the \emph{h-index} of 10.
After reading the papers' abstracts, one was excluded because it was out of the scope of our interest.

\begin{table}[htbp]
\centering
\scalebox{0.75}{
	\small
\begin{tabular}{ c c c }
\hline
\textbf{Year} & \textbf{Records} & \textbf{\% of 34} \\\hline\hline
2020	&	2	&	5.88	\\\hline
2019	&	5	&	14.71	\\\hline
2018	&	6	&	17.65	\\\hline
2017	&	6	&	17.65	\\\hline
2015	&	4	&	11.76	\\\hline
2014	&	3	&	8.82	\\\hline
2012	&	1	&	2.94	\\\hline
2011	&	2	&	5.88	\\\hline
2010	&	2	&	5.88	\\\hline
2009	&	1	&	2.94	\\\hline
2008	&	2	&	5.88	\\\hline
\end{tabular}
}
\caption{Number of papers organized by year.}
\label{table.mountbyyear}
\end{table}

\begin{table*}[htpb] {
	\centering
	%\resizebox{\textwidth}{!}{%
	\scalebox{0.75}{
	\small

\begin{tabular}{c		p{2.5cm}			p{9cm}		p{7cm}		c}
\hline 
%%%%%%%%%%%%%%%%%%%%%%%%%%%%%%%%%%%%%%%%%%
\textbf{\#}	&	\textbf{Author}	&	\textbf{Title}	&	\textbf{Journal}	&	\textbf{Year}	\\\hline
1	&	Dijkman \emph{et al.}	&	Semantics and analysis of business process models in BPMN	&	INFORMATION AND SOFTWARE TECHNOLOGY	&	2008	\\\hline
2	&	Brambilla \emph{et al.}	&	Model-driven design and development of semantic Web service applications	&	ACM TRANSACTIONS ON INTERNET TECHNOLOGY	&	2008	\\\hline
3	&	Thomas \& Fellmann	&	Semantic Process Modeling - Design and Implementation of an Ontology-based Representation of Business Processes	&	BUSINESS \& INFORMATION SYSTEMS ENGINEERING	&	2009	\\\hline
4	&	Lerner \emph{et al.}	&	Exception Handling Patterns for Process Modeling	&	IEEE TRANSACTIONS ON SOFTWARE ENGINEERING	&	2010	\\\hline
5	&	Fernandez Fernandez \emph{et al.}	&	SBPMN - An easier business process modeling notation for business users	&	COMPUTER STANDARDS \& INTERFACES	&	2010	\\\hline
6	&	Wong \&  Gibbons	&	Formalisations and applications of BPMN	&	SCIENCE OF COMPUTER PROGRAMMING	&	2011	\\\hline
7	&	Wong \&  Gibbons	&	Property specifications for workflow modelling	&	SCIENCE OF COMPUTER PROGRAMMING	&	2011	\\\hline
8	&	Perez-Castillo~\emph{et al.}&	A family of case studies on business process mining using MARBLE	&	JOURNAL OF SYSTEMS AND SOFTWARE	&	2012	\\\hline
9	&	Roy \emph{et al.}	&	An Empirical Study of Error Patterns in Industrial Business Process Models	&	IEEE TRANSACTIONS ON SERVICES COMPUTING	&	2014	\\\hline
10	&	Weidlich \emph{et al.}	&	Optimizing Event Pattern Matching Using Business Process Models	&	IEEE TRANSACTIONS ON KNOWLEDGE AND DATA ENGINEERING	&	2014	\\\hline
11	&	Groener \emph{et al.}	&	Validation of user intentions in process orchestration and choreography	&	INFORMATION SYSTEMS	&	2014	\\\hline
12	&	Meyer \emph{et al.}	&	Automating data exchange in process choreographies	&	INFORMATION SYSTEMS	&	2015	\\\hline
13	&	Yousfi \emph{et al.}	&	Introducing decision-aware business processes	&	COMPUTERS IN INDUSTRY	&	2015	\\\hline
14	&	Ramon Stroppi \emph{et al.}	&	Defining the resource perspective in the development of processes-aware information systems	&	INFORMATION AND SOFTWARE TECHNOLOGY	&	2015	\\\hline
15	&	Chandramohan~\emph{et al.} &	Business Process Model for Deriving CIM Profile: A Case Study for Indian Utility	&	IEEE TRANSACTIONS ON POWER SYSTEMS	&	2015	\\\hline
16	&	Kluza \& Nalepa	&	A method for generation and design of business processes with business rules	&	INFORMATION AND SOFTWARE TECHNOLOGY	&	2017	\\\hline
17	&	Poppe \emph{et al.}	&	Design and evaluation of virtual environments mechanisms to support remote collaboration on complex process diagrams	&	INFORMATION SYSTEMS	&	2017	\\\hline
18	&	Khlif \emph{et al.}	&	A methodology for the semantic and structural restructuring of BPMN models	&	BUSINESS PROCESS MANAGEMENT JOURNAL	&	2017	\\\hline
19	&	Gailly \emph{et al.}	&	Recommendation-Based Conceptual Modeling and Ontology Evolution Framework (CMOE plus )	&	BUSINESS \& INFORMATION SYSTEMS ENGINEERING	&	2017	\\\hline
20	&	Baklizky \emph{et al.}	&	Business process point analysis: survey experiments	&	BUSINESS PROCESS MANAGEMENT JOURNAL	&	2017	\\\hline
21	&	Yin \emph{et al.}	&	Service Pattern: An Integrated Business Process Model for Modern Service Industry	&	IEEE TRANSACTIONS ON SERVICES COMPUTING	&	2017	\\\hline
22	&	Yousfi \emph{et al.}	&	Toward uBPMN-Based Patterns for Modeling Ubiquitous Business Processes	&	IEEE TRANSACTIONS ON INDUSTRIAL INFORMATICS	&	2018	\\\hline
23	&	Tang \emph{et al.}	&	Querying Workflow Logs	&	INFORMATION	&	2018	\\\hline
24	&	Mazzola \emph{et al.}	&	Smart Process Optimization and Adaptive Execution with Semantic Services in Cloud Manufacturing	&	INFORMATION	&	2018	\\\hline
25	&	Duran \emph{et al.}	&	Stochastic analysis of BPMN with time in rewriting logic	&	SCIENCE OF COMPUTER PROGRAMMING	&	2018	\\\hline
26	&	Fahland \emph{et al.}	&	Dynamic skipping and Blocking, Dead Path Elimination for Cyclic Workflows, and a Local Semantics for Inclusive Gateways	&	INFORMATION SYSTEMS	&	2018	\\\hline
27	&	Mok 	&	Maximizing control flow concurrency in BPMN workflow models through syntactic means	&	BUSINESS PROCESS MANAGEMENT JOURNAL	&	2018	\\\hline
28	&	Bazhenova \emph{et al.}	&	From BPMN process models to DMN decision models	&	INFORMATION SYSTEMS	&	2019	\\\hline
29	&	Mottola \emph{et al.}	&	makeSense: Simplifying the Integration of Wireless Sensor Networks into Business Processes	&	IEEE TRANSACTIONS ON SOFTWARE ENGINEERING	&	2019	\\\hline
30	&	Muelle \emph{et al.}	&	A practical data-flow verification scheme for business processes	&	INFORMATION SYSTEMS	&	2019	\\\hline
31	&	Sosa-Sanchez \emph{et al.}	&	Aligning Business Processes With the Services Layer Using a Semantic Approach	&	IEEE ACCESS	&	2019	\\\hline
32	&	Combi \emph{et al.}	&	A modular approach to the specification and management of time duration constraints in BPMN	&	INFORMATION SYSTEMS	&	2019	\\\hline
33	&	Xiong \emph{et al.}	&	Detecting Data Flow Errors Across Processes in Business Process Collaboration	&	IEEE ACCESS	&	2020	\\\hline
34	&	Driss \emph{et al.}	&	Servicing Your Requirements: An FCA and RCA-Driven Approach for Semantic Web Services Composition	&	IEEE ACCESS	&	2020	\\\hline
%%%%%%%%%%%%%%%%%%%%%%%%%%%%%%5
\end{tabular}
}
\caption{A List of bibliographic references considered, ordered by year.}
\label{list.publications}
}
\end{table*}

\begin{table*}[htpb]
	\centering
	%\resizebox{\textwidth}{!}{%
	\scalebox{0.55}{
	\small
\begin{tabular}{|p{5cm}|p{0.3cm}|p{0.3cm}|p{0.3cm}|p{0.3cm}|p{0.3cm}|p{0.3cm}|p{0.3cm}|p{0.3cm}|p{0.3cm}|p{0.3cm}|p{0.3cm}|p{0.3cm}|p{0.3cm}|p{0.3cm}|p{0.3cm}|p{0.3cm}|p{0.3cm}|p{0.3cm}|p{0.3cm}|p{0.3cm}|p{0.3cm}|p{0.3cm}|p{0.3cm}|p{0.3cm}|p{0.3cm}|p{0.3cm}|p{0.3cm}|p{0.3cm}|p{0.3cm}|p{0.3cm}|p{0.3cm}|p{0.3cm}|p{0.3cm}|p{0.3cm}|}
\hline
                                
                                         & \rotatebox{90}{Dijkman   \emph{et al.}}  & \rotatebox{90}{Lerner \emph{et al.}} & \rotatebox{90}{Thomas \& Fellmann} & \rotatebox{90}{Wong \&  Gibbons} & \rotatebox{90}{Wong \&  Gibbons} & \rotatebox{90}{Brambilla \emph{et al.}} & \rotatebox{90}{Fernandez Fernandez \emph{et al.}} & \rotatebox{90}{Roy \emph{et al.}} & \rotatebox{90}{Weidlich \emph{et al.}} & \rotatebox{90}{Meyer \emph{et al.}} & \rotatebox{90}{Perez-Castillo~\emph{et al.}} & \rotatebox{90}{Kluza \& Nalepa} & \rotatebox{90}{Yousfi \emph{et al.}} & \rotatebox{90}{Poppe \emph{et al.}} & \rotatebox{90}{Khlif \emph{et al.}} & \rotatebox{90}{Ramon Stroppi \emph{et al.}} & \rotatebox{90}{Yousfi \emph{et al.}} & \rotatebox{90}{Groener \emph{et al.}} & \rotatebox{90}{Bazhenova \emph{et al.}} & \rotatebox{90}{Tang \emph{et al.}} & \rotatebox{90}{Gailly \emph{et al.}} & \rotatebox{90}{Baklizky \emph{et al.}} & \rotatebox{90}{Chandramohan~\emph{et al.}} & \rotatebox{90}{Mottola \emph{et al.}} & \rotatebox{90}{Mazzola \emph{et al.}} & \rotatebox{90}{Muelle \emph{et al.}} & \rotatebox{90}{Sosa-Sanchez \emph{et al.}} & \rotatebox{90}{Duran \emph{et al.}} & \rotatebox{90}{Xiong \emph{et al.}} & \rotatebox{90}{Driss \emph{et al.}} & \rotatebox{90}{Combi \emph{et al.}} & \rotatebox{90}{Fahland \emph{et al.}} & \rotatebox{90}{Mok }       & \rotatebox{90}{Yin \emph{et al.}} \\% \hline
                                                  & \cite{ISI:000259894800007} &  \cite{ISI:000276032300004} & \cite{ISI:000272027400005} & \cite{ISI:000290778300003} & \cite{ISI:000292232900007} & \cite{ISI:000255557600003} & \cite{ISI:000272863000003} & \cite{ISI:000337901500002} & \cite{ISI:000343607500013} & \cite{ISI:000358700800022} & \cite{ISI:000303626300012} & \cite{ISI:000410017900008} & \cite{ISI:000354583400002} & \cite{ISI:000401600000005} & \cite{ISI:000395591400002} & \cite{ISI:000349427200006}  & \cite{ISI:000441446300006} & \cite{ISI:000336110900006} & \cite{ISI:000469906000006} & \cite{ISI:000436147000001} & \cite{ISI:000405735000004} & \cite{ISI:000401004400008} & \cite{ISI:000346734000013} & \cite{ISI:000471686800003} & \cite{ISI:000451310900018} & \cite{ISI:000459839400009} & \cite{ISI:000455865100001} & \cite{ISI:000450385200001} & \cite{ISI:000573790000001} & \cite{ISI:000549809000008} & \cite{ISI:000474502500007} & \cite{ISI:000446949000010} & \cite{ISI:000428168000002} & \cite{ISI:000418057100002} \\ \hline
                                         Model' execution optimization&                                                                                         &                                             &                                             &                                             &                                             &                                             &                                             &                                             &                                             & \CircleSolid                  &                                             & \CircleSolid                  & \CircleSolid                  &                                             &                                             &                                              &                                             &                                             &                                             &                                             &                                             &                                             &                                             &                                             & \CircleSolid                  &                                             &                                             & \CircleSolid                  &                                             &                                             &                                             &                                             & \CircleSolid                  &                                             \\ \hline
Validation                                                                            &                                             &                                             & \CircleSolid                  &                                             &                                             &                                             &                                             &                                             &                                             & \CircleSolid                  &                                             &                                             &                                             &                                             &                                             &                                              &                                             & \CircleSolid                  &                                             &                                             &                                             &                                             &                                             &                                             &                                             &                                             &                                             &                                             & \CircleSolid                  &                                             &                                             &                                             &                                             &                                             \\ \hline
Service architecture focused                                                          &                                             &                                             &                                             &                                             &                                             &                                             &                                             &                                             &                                             &                                             &                                             &                                             &                                             &                                             &                                             &                                              &                                             &                                             &                                             &                                             &                                             &                                             &                                             &                                             & \CircleSolid                  &                                             & \CircleSolid                  &                                             &                                             &                                             &                                             &                                             &                                             & \CircleSolid                  \\ \hline
Querying                                                                              &                                             &                                             & \CircleSolid                  &                                             &                                             &                                             &                                             &                                             &                                             &                                             &                                             &                                             &                                             &                                             &                                             &                                              &                                             &                                             &                                             & \CircleSolid                  &                                             &                                             &                                             &                                             &                                             &                                             &                                             &                                             &                                             &                                             &                                             &                                             &                                             &                                             \\ \hline
Event Monitoring                                                                     &                                             &                                             &                                             &                                             &                                             &                                             &                                             &                                             & \CircleSolid                  &                                             &                                             &                                             &                                             &                                             &                                             &                                              &                                             &                                             &                                             &                                             &                                             &                                             &                                             &                                             &                                             &                                             &                                             &                                             &                                             &                                             &                                             &                                             &                                             &                                             \\ \hline
Business process re-engineering                                                        &                                             &                                             &                                             &                                             &                                             &                                             &                                             &                                             &                                             &                                             &                                             &                                             &                                             &                                             & \CircleSolid                  &                                              &                                             &                                             &                                             &                                             &                                             &                                             &                                             &                                             &                                             &                                             &                                             &                                             &                                             &                                             &                                             &                                             &                                             &                                             \\ \hline
Source code generation                                                                &                                             &                                             &                                             &                                             &                                             &                                             &                                             &                                             &                                             &                                             &                                             &                                             &                                             &                                             &                                             &                                              &                                             &                                             &                                             &                                             &                                             &                                             &                                             & \CircleSolid                  &                                             &                                             &                                             &                                             &                                             &                                             &                                             &                                             &                                             &                                             \\ \hline
\end{tabular}
}
\caption{List of applications that are referred from related work.}
\label{tab.applications}
\end{table*}

Table~\ref{tab.applications} shows a list of applications that are understood from this set of related papers, and Table~\ref{tab.modellinglanguages} presents a list of the modeling languages that are referred. We remark that BPMN 2.0 is broadly used what reinforces the importance of finding semantic pattern to be used by practitioners and industry. Petri nets are used often to formalize their proposals. The remaining modeling languages used are residual.
Next, the papers are divided in two groups: semantic- and syntactic-based approaches.

\begin{table*}[htpb]
	\centering
	%\resizebox{\textwidth}{!}{%
	\scalebox{0.55}{
	\small
\begin{tabular}{|p{4cm}|p{0.3cm}|p{0.3cm}|p{0.3cm}|p{0.3cm}|p{0.3cm}|p{0.3cm}|p{0.3cm}|p{0.3cm}|p{0.3cm}|p{0.3cm}|p{0.3cm}|p{0.3cm}|p{0.3cm}|p{0.3cm}|p{0.3cm}|p{0.3cm}|p{0.3cm}|p{0.3cm}|p{0.3cm}|p{0.3cm}|p{0.3cm}|p{0.3cm}|p{0.3cm}|p{0.3cm}|p{0.3cm}|p{0.3cm}|p{0.3cm}|p{0.3cm}|p{0.3cm}|p{0.3cm}|p{0.3cm}|p{0.3cm}|p{0.3cm}|p{0.3cm}|}
\hline
                          & \rotatebox{90}{Dijkman \emph{et   al.}} & \rotatebox{90}{Lerner \emph{et   al.}} & \rotatebox{90}{Thomas \&   Fellmann} & \rotatebox{90}{Wong \&  Gibbons} & \rotatebox{90}{Wong \&  Gibbons} & \rotatebox{90}{Brambilla \emph{et   al.}} & \rotatebox{90}{Fernandez Fernandez~\emph{et al.}} & \rotatebox{90}{Roy \emph{et al.}} & \rotatebox{90}{Weidlich \emph{et   al.}} & \rotatebox{90}{Meyer \emph{et   al.}} & \rotatebox{90}{Perez-Castillo~\emph{et al.}} & \rotatebox{90}{Kluza \&   Nalepa} & \rotatebox{90}{Yousfi \emph{et   al.}} & \rotatebox{90}{Poppe \emph{et   al.}} & \rotatebox{90}{Khlif \emph{et   al.}} & \rotatebox{90}{Ramon Stroppi   \emph{et al.}} & \rotatebox{90}{Yousfi \emph{et   al.}} & \rotatebox{90}{Groener \emph{et   al.}} & \rotatebox{90}{Bazhenova \emph{et   al.}} & \rotatebox{90}{Tang \emph{et al.}} & \rotatebox{90}{Gailly \emph{et   al.}} & \rotatebox{90}{Baklizky \emph{et   al.}} & \rotatebox{90}{Chandramohan~\emph{et al.}} & \rotatebox{90}{Mottola \emph{et   al.}} & \rotatebox{90}{Mazzola \emph{et   al.}} & \rotatebox{90}{Muelle \emph{et   al.}} & \rotatebox{90}{Sosa-Sanchez   \emph{et al.}} & \rotatebox{90}{Duran \emph{et   al.}} & \rotatebox{90}{Xiong \emph{et   al.}} & \rotatebox{90}{Driss \emph{et   al.}} & \rotatebox{90}{Combi \emph{et   al.}} & \rotatebox{90}{Fahland \emph{et   al.}} & \rotatebox{90}{Mok }       & \rotatebox{90}{Yin \emph{et   al.}} \\ 
                          & \cite{ISI:000259894800007}              & \cite{ISI:000276032300004}             & \cite{ISI:000272027400005}           & \cite{ISI:000290778300003}       & \cite{ISI:000292232900007}       & \cite{ISI:000255557600003}                & \cite{ISI:000272863000003}                                  & \cite{ISI:000337901500002}        & \cite{ISI:000343607500013}               & \cite{ISI:000358700800022}            & \cite{ISI:000303626300012}                     & \cite{ISI:000410017900008}        & \cite{ISI:000354583400002}             & \cite{ISI:000401600000005}            & \cite{ISI:000395591400002}            & \cite{ISI:000349427200006}                    & \cite{ISI:000441446300006}             & \cite{ISI:000336110900006}              & \cite{ISI:000469906000006}                & \cite{ISI:000436147000001}         & \cite{ISI:000405735000004}             & \cite{ISI:000401004400008}               & \cite{ISI:000346734000013}                   & \cite{ISI:000471686800003}              & \cite{ISI:000451310900018}              & \cite{ISI:000459839400009}             & \cite{ISI:000455865100001}                   & \cite{ISI:000450385200001}            & \cite{ISI:000573790000001}            & \cite{ISI:000549809000008}            & \cite{ISI:000474502500007}            & \cite{ISI:000446949000010}              & \cite{ISI:000428168000002} & \cite{ISI:000418057100002}          \\ \hline
BPMN 2.0                  & \CircleSolid                            & \CircleSolid                           & \CircleSolid                         & \CircleSolid                     & \CircleSolid                     & \CircleSolid                              & \CircleSolid                                                & \CircleSolid                      & \CircleSolid                             & \CircleSolid                          & \CircleSolid                                   & \CircleSolid                      & \CircleSolid                           & \CircleSolid                          & \CircleSolid                          & \CircleSolid                                  & \CircleSolid                           & \CircleSolid                            & \CircleSolid                              & \CircleSolid                       & \CircleSolid                           & \CircleSolid                             & \CircleSolid                                 & \CircleSolid                            & \CircleSolid                            & \CircleSolid                           & \CircleSolid                                 & \CircleSolid                          & \CircleSolid                          & \CircleSolid                          & \CircleSolid                          & \CircleSolid                            & \CircleSolid               & \CircleSolid                        \\ \hline
Petri nets                & \CircleSolid                            &                                        &                                      & \CircleSolid                     &                                  &                                           &                                                             & \CircleSolid                      & \CircleSolid                             & \CircleSolid                          &                                                &                                   &                                        &                                       &                                       &                                               &                                        &                                         &                                           &                                    &                                        &                                          &                                              &                                         &                                         & \CircleSolid                           &                                              &                                       &                                       &                                       & \CircleSolid                          & \CircleSolid                            &                            &                                     \\ \hline
UML 2.0 Activity Diagrams &                                         & \CircleSolid                           &                                      &                                  &                                  &                                           &                                                             &                                   &                                          &                                       &                                                &                                   &                                        &                                       &                                       &                                               &                                        &                                         &                                           &                                    &                                        &                                          &                                              &                                         &                                         &                                        &                                              &                                       &                                       &                                       &                                       &                                         &                            &                                     \\ \hline
Little-JIL                &                                         & \CircleSolid                           &                                      &                                  &                                  &                                           &                                                             &                                   &                                          &                                       &                                                &                                   &                                        &                                       &                                       &                                               &                                        &                                         &                                           &                                    &                                        &                                          &                                              &                                         &                                         &                                        &                                              &                                       &                                       &                                       &                                       &                                         &                            &                                     \\ \hline
UML                       &                                         &                                        &                                      &                                  &                                  &                                           &                                                             &                                   &                                          &                                       &                                                &                                   &                                        &                                       &                                       &                                               &                                        &                                         &                                           &                                    &                                        &                                          & \CircleSolid                                 &                                         &                                         &                                        &                                              &                                       &                                       &                                       &                                       &                                         &                            &                                     \\ \hline
WebML                     &                                         &                                        &                                      &                                  &                                  & \CircleSolid                              &                                                             &                                   &                                          &                                       &                                                &                                   &                                        &                                       &                                       &                                               &                                        &                                         &                                           &                                    &                                        &                                          &                                              &                                         &                                         &                                        &                                              &                                       &                                       &                                       &                                       &                                         &                            &                                     \\ \hline
EPC                       &                                         &                                        & \CircleSolid                         &                                  &                                  &                                           &                                                             &                                   &                                          &                                       &                                                &                                   &                                        &                                       &                                       &                                               &                                        &                                         &                                           &                                    &                                        &                                          &                                              &                                         &                                         &                                        &                                              &                                       &                                       &                                       &                                       &                                         &                            &                                     \\ \hline
Goal model                &                                         &                                        &                                      &                                  &                                  &                                           &                                                             &                                   &                                          &                                       &                                                &                                   &                                        &                                       &                                       &                                               &                                        & \CircleSolid                            &                                           &                                    &                                        &                                          &                                              &                                         &                                         &                                        &                                              &                                       &                                       &                                       &                                       &                                         &                            &                                     \\ \hline

\end{tabular}
}
\caption{List of modeling languages that are referred from related work.}
\label{tab.modellinglanguages}
\end{table*}

\subsection{Semantic-based approaches}

An overview is depicted in Table~\ref{tab.semantic}. Concerning ontologies, only two papers using UFO~\cite{guizzardi2015towards} and OWL~\cite{owl} are directly related (\emph{cf.} Table~\ref{tab.ontologies}).

\begin{table}[htpb]
	\centering
	%\resizebox{\textwidth}{!}{%
	\scalebox{0.55}{
	\small

\begin{tabular}{|p{3cm}|p{0.3cm}|p{0.3cm}|p{0.3cm}|p{0.3cm}|p{0.3cm}|p{0.3cm}|p{0.3cm}|p{0.3cm}|p{0.3cm}|p{0.3cm}|p{0.3cm}|p{0.3cm}|p{0.3cm}|p{0.3cm}|p{0.3cm}|p{0.3cm}|}
\hline
                                        & \rotatebox{90}{Dijkman \emph{et al.}} & \rotatebox{90}{Lerner \emph{et al.}} & \rotatebox{90}{Thomas \& Fellmann} & \rotatebox{90}{Wong \&  Gibbons} & \rotatebox{90}{Wong \&  Gibbons} & \rotatebox{90}{Roy \emph{et al.}} & \rotatebox{90}{Weidlich \emph{et al.}} & \rotatebox{90}{Meyer \emph{et al.}} & \rotatebox{90}{Perez-Castillo~\emph{et al.}} & \rotatebox{90}{Poppe \emph{et al.}} & \rotatebox{90}{Khlif \emph{et al.}} & \rotatebox{90}{Tang \emph{et al.}} & \rotatebox{90}{Mazzola \emph{et al.}} & \rotatebox{90}{Muelle \emph{et al.}} & \rotatebox{90}{Sosa-Sanchez \emph{et al.}} & \rotatebox{90}{Xiong \emph{et al.}} \\ 
                                        & \cite{ISI:000259894800007}            & \cite{ISI:000276032300004}           & \cite{ISI:000272027400005}         & \cite{ISI:000290778300003}       & \cite{ISI:000292232900007}       & \cite{ISI:000337901500002}        & \cite{ISI:000343607500013}             & \cite{ISI:000358700800022}          & \cite{ISI:000303626300012}                     & \cite{ISI:000401600000005}          & \cite{ISI:000395591400002}          & \cite{ISI:000436147000001}         & \cite{ISI:000451310900018}            & \cite{ISI:000459839400009}           & \cite{ISI:000455865100001}                 & \cite{ISI:000573790000001}          \\ \hline
Semantic analysis of BPMN model         & \CircleSolid                          &                                      &                                    & \CircleSolid                     & \CircleSolid                     &                                   & \CircleSolid                           & \CircleSolid                        &                                                & \CircleSolid                        &                                     &                                    &                                       & \CircleSolid                         & \CircleSolid                               &                                     \\ \hline
Incident or Error or exception patterns &                                       & \CircleSolid                         &                                    &                                  &                                  & \CircleSolid                      &                                        &                                     &                                                &                                     &                                     & \CircleSolid                       &                                       &                                      &                                            &                                     \\ \hline
Communication/data- flow focus          &                                       &                                      &                                    &                                  &                                  &                                   &                                        & \CircleSolid                        &                                                &                                     &                                     &                                    &                                       & \CircleSolid                         &                                            & \CircleSolid                        \\ \hline
Semantic annotations                    &                                       &                                      & \CircleSolid                       &                                  &                                  &                                   &                                        &                                     &                                                &                                     &                                     &                                    & \CircleSolid                          &                                      &                                            &                                     \\ \hline
Business experts perspective            &                                       &                                      &                                    &                                  &                                  &                                   &                                        &                                     &                                                &                                     &                                     & \CircleSolid                       &                                       &                                      &                                            &                                     \\ \hline
Retrieve business process knowlegde     &                                       &                                      &                                    &                                  &                                  &                                   &                                        &                                     & \CircleSolid                                   &                                     &                                     &                                    &                                       &                                      &                                            &                                     \\ \hline
Heuristic definition                    &                                       &                                      &                                    &                                  &                                  &                                   &                                        &                                     &                                                &                                     & \CircleSolid                        &                                    &                                       &                                      &                                            &                                     \\ \hline
\end{tabular}

}
\caption{List of semantic-based solutions in the related work.}
\label{tab.semantic}
\end{table}

In~\cite{ISI:000259894800007} the authors state that the mix of constructs found in BPMN makes it possible to create models with semantic errors. Such errors could be propagated to the later phases of systems development which are referred among the most costly and hardest to correct. The paper proposes a mapping between a large subset of BPMN to petri nets to define a formal semantics. A tool has been developed and tested with 13 models available from BPMN Wikipedia and detected errors in 3 models. 
\cite{ISI:000459839400009} proposes a data-flow verification scheme for business processes. To reduce the state-space of a model, a high-level representation based in the notion of relevance.
Both~\cite{ISI:000259894800007} and~\cite{ISI:000459839400009} could be used to improve the correctness of user defined happy-path models.

Our approach does not provides a formal way to prevent errors, but do lead to a reduction of errors because users tend to focus on the happy path, which due to its simplicity yields fewer errors. The more complex handling of declinations and revocations are generated from the patterns and thus are error free.
 
The alignment between business processes and services layer is studied in~\cite{ISI:000455865100001}. A solution based on the semantic extraction of the BPMN elements is tested. We consider that this alignment could be better achieved if the BPMN models include the correct, and complete, semantics from business, as offered by our proposal. Moreover, semantic cannot be directly extracted from BPMN elements, but rather from a business definition as presented in section~\ref{omegaA.validation} of our paper.

Authors in~\cite{ISI:000395591400002} propose a solution to restructure the structure of a business process model using semantic information based on data from a social network, hierarchical clustering and graph optimization. Nonetheless the importance that AI solutions have today, our solution offers a more robust output since it fully describe how a network of business processes should be designed, even if some parts of it are, at a specific instant in time, considered as implicit. The outcome of this paper always depends on the quality delivered by the heuristic and on the set of rules defined.
In the scope of process-aware information systems,~\cite{ISI:000349427200006} propose a set of patterns related with resources with an workflow management system. This approach is narrower than our which allows the application to all the context of a business transaction.

~\cite{ISI:000272027400005} proposes the formalization of each model element using ontology to cope with misinterpretations. Authors state that this approach allows the model querying and its validation. This approach allows a broader application to models, but leaves the details open to the ontology definition. Therefore, it is less focused in the specific semantics of business processes dynamics when compared with our proposal.

\cite{ISI:000436147000001} uses an enterprise' incident patterns to propose a querying solution based on business experts perspectives instead of a classical approach of ETL/data warehousing solution. Similarly with our proposal, we emphasize the need to find patterns to empower the knowledge from business directly to the business processes. 

The work of~\cite{ISI:000549809000008} use ontological description and BPMN to capture the semantic modeling of requirements. However, the connection between both is not detailed.

To optimize the model' efficiency the proposal of~\cite{ISI:000451310900018} grounds on formalized domain knowledge and structured service wrapped with semantic annotation that is considered a BPMN enrichment. 

Also corroborating the need to strengthen the knowledge of business processes and its importance, in~\cite{ISI:000343607500013} authors uses knowledge of business processes to optimize the event pattern matching used in BAM or SLAs platforms. Despite the different application it is worth noticing that giving meaning to business process allows a more precise (with less computational effort) operational monitoring. 

\cite{ISI:000255557600003} state that business processes models and web engineering models are enough to support the semi-automatic semantic extractions and thus produce rich web applications. 
Nevertheless, models need to be complete and therefore our approach also applies in to improve their proposal.

In~\cite{ISI:000276032300004} authors distinguish between the normative desirable behaviour that is sometimes referred to as the ``happy path" with the deviations from that happy path as exception handling. A set of exceptions patterns are proposed in the paper and its existence is evaluated.﻿ From the models considered, results suggest that the proposed exception handling patterns were instances of the patterns from the catalogue. 

In addition, the authors of~\cite{ISI:000346734000013} that use BPMN to derive a common information model. We note that these approaches are only possible if the supporting BPMN models are semantically correct. Therefore, we see our work as an enabler to validate the semantic correctness.

The work in~\cite{ISI:000303626300012} uses process mining to retrieve legacy business processes using a static analysis of source code. The core idea is to preserve the business knowledge. This study corroborates the importance of the semantics of business, yet do not limit multiple interpretations of the same business processes.  

~\cite{ISI:000401600000005} studies the problem of multiple interpretation between stakeholders in the specific context of remote collaborations. Here the problem is more complex and drives to mode coordination problems. This study corroborates our initial problem statement and reinforces it when remote environments are imposed, \emph{e.g.}, off-shore and near-shore development.

In~\cite{ISI:000573790000001} authors point to existing limitations in the communications between collaboration specifications on business process collaborations. To solve this problem, a data flow-error detecting based on a complete data model of all BPMN elements is required. Our proposal inherits the DEMO communication pattern avoiding modeling errors.

\cite{ISI:000354583400002,ISI:000410017900008} introduce the concern of business rules along with business process design. The concept of Decision-aware Business Processes is used as a refinement of the concept of decision having the goal of optimizing the monitoring time of each decision.

Authors in~\cite{ISI:000336110900006} use description logic and automated reasoners to validate if the goals that are extracted from requirements are designed in the activities within business processes. Two inconsistencies are identified: orchestration and choreography. The solution is able to validate, but do not specify how to produce the models.

%Table~\ref{tab.ontologies} presents the list of ontologies that are referenced in the related work.

\begin{table}[htpb]
	\centering
	\scalebox{0.55}{
	\small

\begin{tabular}{|p{1cm}|p{2cm}|p{2cm}|p{2cm}|p{2cm}|p{2cm}|}
\hline
%    & \rotatebox{90}{Thomas \&   Fellmann} & \rotatebox{90}{Gailly \emph{et   al.}} & \rotatebox{90}{Mazzola \emph{et   al.}} & \rotatebox{90}{Driss \emph{et   al.}} \\ 
    & Thomas \&   Fellmann \cite{ISI:000272027400005}& Gailly \emph{et   al.} \cite{ISI:000405735000004} & Mazzola \emph{et   al.}  \cite{ISI:000451310900018}& Driss \emph{et   al.} \cite{ISI:000549809000008} \\ \hline
OWL~\cite{owl} & \CircleSolid                         &                                        & \CircleSolid                            & \CircleSolid                          \\ \hline
UFO~\cite{guizzardi2015towards} &                                      & \CircleSolid &                                                &                                       \\ \hline
\end{tabular}

}
\caption{List of ontologies referenced in related work.}
\label{tab.ontologies}
\end{table}

\subsection{Syntactic-based approaches}

A summary of the syntactic-based contributions is summarized in Table~\ref{tab.syntactic}.

\begin{table}[htpb]
	\centering
	%\resizebox{\textwidth}{!}{%
	\scalebox{0.55}{
	\small
\begin{tabular}{|p{3cm}|p{0.3cm}|p{0.3cm}|p{0.3cm}|p{0.3cm}|p{0.3cm}|p{0.3cm}|p{0.3cm}|p{0.3cm}|p{0.3cm}|p{0.3cm}|p{0.3cm}|p{0.3cm}|p{0.3cm}|p{0.3cm}|p{0.3cm}|p{0.3cm}|}
\hline
                                   & \rotatebox{90}{Wong \&  Gibbons} & \rotatebox{90}{Wong \&  Gibbons} & \rotatebox{90}{Fernandez Fernandez~\emph{et al.}} & \rotatebox{90}{Roy \emph{et al.}} & \rotatebox{90}{Perez-Castillo~\emph{et al.}} & \rotatebox{90}{Ramon Stroppi \emph{et al.}} & \rotatebox{90}{Yousfi \emph{et al.}} & \rotatebox{90}{Bazhenova \emph{et al.}} & \rotatebox{90}{Gailly \emph{et al.}} & \rotatebox{90}{Baklizky \emph{et al.}} & \rotatebox{90}{Mottola \emph{et al.}} & \rotatebox{90}{Sosa-Sanchez \emph{et al.}} & \rotatebox{90}{Duran \emph{et al.}} & \rotatebox{90}{Combi \emph{et al.}} & \rotatebox{90}{Fahland \emph{et al.}} & \rotatebox{90}{Mok }       \\ 
                                   & \cite{ISI:000290778300003}       & \cite{ISI:000292232900007}       & \cite{ISI:000272863000003}                                  & \cite{ISI:000337901500002}        & \cite{ISI:000303626300012}                     & \cite{ISI:000349427200006}                  & \cite{ISI:000441446300006}           & \cite{ISI:000469906000006}              & \cite{ISI:000405735000004}           & \cite{ISI:000401004400008}             & \cite{ISI:000471686800003}            & \cite{ISI:000455865100001}                 & \cite{ISI:000450385200001}          & \cite{ISI:000474502500007}          & \cite{ISI:000446949000010}            & \cite{ISI:000428168000002} \\ \hline
Syntactic analysis                 &                                  &                                  &                                                             & \CircleSolid                      & \CircleSolid                                   &                                             &                                      & \CircleSolid                            & \CircleSolid                         &                                        & \CircleSolid                          & \CircleSolid                               & \CircleSolid                        & \CircleSolid                        & \CircleSolid                          & \CircleSolid               \\ \hline
BPMN notation enrichment           &                                  &                                  & \CircleSolid                                                &                                   &                                                &                                             & \CircleSolid                         &                                         &                                      &                                        & \CircleSolid                          &                                            & \CircleSolid                        &                                     &                                       &                            \\ \hline
Process algebra CSP                & \CircleSolid                     & \CircleSolid                     &                                                             &                                   &                                                &                                             &                                      &                                         &                                      &                                        &                                       &                                            &                                     &                                     &                                       &                            \\ \hline
Property specification language PL & \CircleSolid                     & \CircleSolid                     &                                                             &                                   &                                                &                                             &                                      &                                         &                                      &                                        &                                       &                                            &                                     &                                     &                                       &                            \\ \hline
Function points analysis           &                                  &                                  &                                                             &                                   &                                                &                                             &                                      &                                         &                                      & \CircleSolid                           &                                       &                                            &                                     &                                     &                                       &                            \\ \hline
Workflow resource perspective      &                                  &                                  &                                                             &                                   &                                                & \CircleSolid                                &                                      &                                         &                                      &                                        &                                       &                                            &                                     &                                     &                                       &                            \\ \hline
Time-constraint enforcement        &                                  &                                  &                                                             &                                   &                                                &                                             &                                      &                                         &                                      &                                        &                                       &                                            &                                     & \CircleSolid                        &                                       &                            \\ \hline
\end{tabular}
}
\caption{List of syntactic-based solutions in the related work.}
\label{tab.syntactic}
\end{table}

Authors of~\cite{ISI:000290778300003,ISI:000292232900007} defines a formal syntactic approach using process algebra CSP, and property specification language PL, applied to a subset of BPMN specification to ensure precise business specification and to assist developers in implementing the correct business processes. Yet, the formalism, and subsequent patterns, are done at the BPMN constructs levels, instead of our approach that considers the business transaction as the basic construct. In opposite, other authors extract semantic from BPMN models to produce artifacts, \emph{e.g.}, \cite{ISI:000471686800003} propose a BPMN extension to model the context of wireless sensor networks. The goal is to facilitate the generation of source-code from the business models for this specific domain.

~\cite{ISI:000337901500002} studies the error patterns that are induced in industrial business processes obtained from syntactic analysis. Though the semantic has not been considered, it is showed by empirical evidences that an high percentage of errors exist implying a need to improve business process design.

Authors in~\cite{ISI:000272863000003} propose a new notation for BPMN alleging that BPMN notation is misunderstood by business stakeholders. And, that factor is considered as an impediment to the correct expressiveness of business processes semantic.
Despite the fact that we agree with such claim, our approach do allow business stakeholders to understand BPMN models because they only need to design the happy path.
Also,~\cite{ISI:000441446300006} extends BPMN with more notations to deal with ubiquitous computing technologies alleging that BPMN do not offer support for this domain.  

\cite{ISI:000446949000010} is focused on the syntactic analysis of the process model to solve the open problem to define dead-path-elimination for cyclic workflows. This solution is directly applicable to workflow engines. Those recommended patterns in~\cite{ISI:000446949000010} could be beneficial to assess whether the business processes generated by our solution, in specific when a huge network of transactions exists, are syntactically correct in BPMN.

\cite{ISI:000405735000004} corroborates the different interpretations of a given business process that could be conflicting and propose a semantic repository to inter-operate models. However, the repository is based on the syntax of BPMN and not on the semantic of business process. This approach could be beneficial if applied to our semantic pattern to offer a recommendation between multiple business processes options.

The work of~\cite{ISI:000401004400008} applies the function-points analysis solution to BPMN business processes to evaluate them. The syntax of each business process is assessed directly by experimental studies. This is a different approach from ours, where we aim to automate the design of business process with a given semantic. 

In~\cite{ISI:000474502500007} authors present a solution to control the time during the life-cycle of a BPMN activity, named as a duration-aware model. This work assures time constraints which is absent in BPMN specification. The integration of time constraints with our proposal allows the fine-grained control of DEMO acts within the execution of complex networks of transactions.

Authors in~\cite{ISI:000418057100002} introduce the business processes patterns from the point view of resource and data with formalized description. In comparison with our proposal, the patterns are presented but its implementation is not further detailed.

The paper~\cite{ISI:000428168000002} refers that the semantics of BPMN control flow is enforced by user. Starting on this premise a syntactic analysis is performed to optimize the model' execution efficiency concerning the maximization of concurrency. In relation with our proposal, the DEMO patterns are used as the basis for a business process. Then, the specific activities that  are performed in each transaction act may benefit from optimization.

The work in~\cite{ISI:000358700800022} propose an enrichment to BPMN choreography in a similar way with our proposal. However, only a request/response pattern is used and do not specify how to enforce in BPMN collaborations. Our proposal is more complete due to the DEMO complete theory that underlies the implementation.

Similarly with~\cite{ISI:000354583400002}, the paper in~\cite{ISI:000469906000006} studies the problem of decoding the decisions designed in a BPMN model. A set of patterns is proposed to capture the possible representations of data in a BPMN model. 

In~\cite{ISI:000450385200001} is stated that literature appears to have few support for understanding the quantitative aspects in the business process design. Therefore, authors propose an executable specification of BPMN using a stochastic approach that analyses time and branching constructs. This approach could beneficial to analyze the results offered by our proposal.

\section{Conclusions and future work}
\label{omegaA.conclusions}

This paper proposes a framework that maps the DEMO complete pattern onto BPMN oferring a semantified BPMN pattern to model business processes.
The pattern encompasses the ability to express both the communication and the production acts, as well as, the ability to systematically prescribe revocations.
Moreover, the pattern can be combined in a network offering the ability to design and implement within a controlled managed complexity environment. 

The main contributions of this paper are
\emph{(i)} the increase in BPMN models completeness by adding patterns, with a reduced, and thus managed, increase in complexity. This has major impact on executable models maintenance, reuse and testing; 
\emph{(ii)} the standardization of communications between pools using the Enterprise Ontology body of knowledge;
\emph{(iii)} the application to two proof-of-concepts;
\emph{(iv)} an extensive literature review to assess our work with the available knowledge in this domain; and
\emph{(v)} a tool to generate the mapping, whose besides model analysis, the BPMN file output could be used to execute BPMN processes in an engine.

From here, several endeavours for future work follows. 
%Firstly, the development of a software artifact to produce the mappings between DEMO and BPMN automatically. The inputs are: list of actors, list of transactions and the list of dependencies between transactions. The output is a BPMN model based in the available BPMN XSD definition~\cite{bpmnXSD}.
Firstly, the propagation of revocations within a network of transactions.
Secondly, develop the ability to produce views of the BPMN model produced. The idea is to filter complexity, \emph{e.g.}, allowing the definition of a view vector, for each transaction, specifying which acts or revocations are required.  
Thirdly, the development of a tool to generate the test cases that support the verification and validation of the BPMN model.

\section*{Acknowledgements}
The authors state that this work was supported by the European Commission program H2020 under the grant agreement 822404 (project QualiChain) and by national funds through Funda\c{c}\~ao para a Ci\^encia e a Tecnologia (FCT) with reference UIDB/50021/2020 (INESC-ID).

\bibliography{reference}

\end{document}